\documentclass[a4paper,11pt]{article}

\usepackage{jheppub} 

\usepackage[T1]{fontenc} 
\usepackage{slashed}
\usepackage{cancel}
\usepackage{lipsum}
\usepackage{graphicx}
\usepackage{dcolumn}
\usepackage{bm}
\usepackage[dvipsnames,usenames]{xcolor}
\usepackage{amsmath}
\usepackage{amssymb} 
\usepackage{hyperref}
\usepackage{bbold}
\usepackage{algpseudocode}
\usepackage{mathtools}
\usepackage{multirow}
\usepackage{caption}
\usepackage{subcaption}
\usepackage[compat=1.1.0]{tikz-feynman}

\DeclareMathOperator\PolyLog{Li_{2}}

\newcommand{\nc}{\newcommand}
\newcommand{\rc}{\renewcommand}
\nc{\bfx}{{\bf x}}
\nc{\bfy}{{\bf y}}
\nc{\bfz}{{\bf z}}
\nc{\bfxh}{{\bf \hat{x}}}
\nc{\bfyh}{{\bf \hat{y}}}
\nc{\bfzh}{{\bf \hat{z}}}
\nc{\bfj}{{\bf j}}
\nc{\bfr}{{\bf r}}
\nc{\bfR}{{\bf R}}
\nc{\bfk}{{\bf k}}
\nc{\bfq}{{\bf q}}
\nc{\bfp}{{\bf p}}
\nc{\bfv}{{\bf v}}
\nc{\bfvh}{{\bf \hat{v}}}
\nc{\bfqh}{{\bf \hat{q}}}

\nc{\mel}[3]{\langle #1 | #2 | #3 \rangle}
\nc{\rme}[1]{\lange || #1 || \rangle}
\nc{\nuc}[2]{^{#1}{\rm #2}}
\nc{\me}[1]{\mel{\nuc{6}{Li}}{#1}{\nuc{6}{He}}}
\nc{\mee}[1]{\mel{\nuc{6}{Li},10}{#1}{\nuc{6}{He},00}}
\nc{\ord}[1]{\mathcal{O}(#1)}
\rc{\Re}{{\rm Re}}
\rc{\Im}{{\rm Im}}

\title{\boldmath
Radiative corrections to superallowed beta decays at $\mathcal O(\alpha^2 Z)$.}

\author[a,b,c]{\`O. L. Crosas}
\author[c]{E. Mereghetti}

\affiliation[a]{Physik-Institut, Universität Zürich, Winterthurerstrasse 190, CH-8057 Zürich, Switzerland}
\affiliation[b]{PSI Center for Neutron and Muon Sciences, 5232 Villigen PSI, Switzerland}
\affiliation[c]{Theoretical Division, Los Alamos National Laboratory, Los Alamos, NM 87545, USA }

\emailAdd{oscar.laracrosas@physik.uzh.ch}
\emailAdd{emereghetti@lanl.gov}

\abstract{
We compute $\mathcal O(\alpha^2 Z)$ radiative corrections to superallowed $\beta$ decays with a heavy-particle effective field theory that systematically describes the interactions of low-energy ultrasoft photons with nuclei. We calculate two-loop virtual and one-loop real-virtual amplitudes by reducing the Feynman integrals to a set of master integrals, which we solve analytically using a variety of techniques. These techniques can be applied to other phenomenologically interesting observables. The ultrasoft corrections can then be combined with contributions arising from the exchange of potential photons to obtain the complete $\mathcal O(\alpha^2 Z)$ correction to the decay rate, with resummation of large logarithms of the electron energy times the nuclear radius. 
We find that $\mathcal O(\alpha^2 Z)$ ultrasoft loops induce a relative correction to the decay rate that ranges from  $0.7 \cdot 10^{-3}$ in the decay of $^{10}$C to  
$3.6 \cdot 10^{-3}$ in the decay of $^{54}$Co, and will thus impact the extraction of $V_{ud}$ at the permille level. We show  that the inclusion of these corrections reduces the residual renormalization scale dependence of the decay rate to a negligible level, making missing ultrasoft perturbative corrections  
a subdominant source of theoretical uncertainty.
}

\begin{document} 
\maketitle
\flushbottom

\section{Introduction}

Tests of the unitarity of the 
Cabibbo--Kobayashi--Maskawa (CKM) quark mixing matrix~\cite{Cabibbo:1963yz,Kobayashi:1973fv} are a powerful tool to probe physics beyond the Standard Model at scales comparable to those directly accessible at high energy colliders~\cite{Czarnecki:2004cw,Towner:2010zz,ParticleDataGroup:2024cfk, Hocker:2001xe,UTfit:2005ras,Seng:2018yzq,Hardy:2020qwl,Cirigliano:2022yyo}.
Superallowed $\beta$ decays 
currently provide the most 
precise determination of the $V_{ud}$ element of the CKM matrix, with a relative uncertainty of $\delta V_{ud}/V_{ud} = 3 \cdot 10^{-4}$.
This level of precision can be reached because the conservation of the weak vector current ensures that the leading Fermi nuclear matrix element in $0^+ \rightarrow 0^+$ transitions is simply determined by a Clebsch-Gordan coefficient, 
$M_F^{(0)} =\sqrt{2}$,
and that the first corrections are suppressed by powers of the electromagnetic coupling $\alpha$. 
The calculation of electromagnetic corrections to $\beta$ decay has a long history~\cite{Kinoshita:1958ru,Sirlin:1967zz,Sirlin:1967zza,Abers:1968zz,Jaus:1970tah,Sirlin:1977sv,Sirlin:1981ie,Wilkinson:1982hu,Sirlin:1986cc,Towner:1992xm,
Wilkinson:1993hxz,Wilkinson:1993fva,Czarnecki:2004cw,Marciano:2005ec,Towner:2010zz,Seng:2018yzq,Seng:2018qru,Gorchtein:2018fxl,Seng:2022cnq,Ma:2023kfr,Seng:2023cvt,Hill:2023acw,Cirigliano:2023fnz,Cirigliano:2024rfk,Cirigliano:2024msg,Borah:2024ghn,Gennari:2024sbn},
whose development even predates the formulation of the Standard Model. 
The multiscale nature of nuclear $\beta$ decays implies that electromagnetic corrections arise from photon exchanges at different energy scales: the electroweak scale~\cite{Sirlin:1981ie,Hill:2019xqk,Dekens:2019ept}, the scale at which QCD becomes nonperturbative~\cite{Marciano:2005ec,Seng:2018yzq,Seng:2018qru}, the typical scale of nuclear physics~\cite{Towner:1992xm}, down to the low-energy scales of the reactions' $\mathcal Q$-values and of the electron mass~\cite{Sirlin:1967zza}.
The separation between these disparate scales can be leveraged to frame the problem using a tower of Effective Field Theories (EFTs)~\cite{Hill:2023acw,Cirigliano:2023fnz,Cirigliano:2024rfk,Cirigliano:2024msg}, which allows one to isolate the contributions of different photon modes into functions that depend on a single scale~\cite{Cirigliano:2024msg}.
The EFT formalism and the mapping between objects defined in the EFT and the traditional corrections, used for example in  the most recent extraction of $V_{ud}$~\cite{Hardy:2020qwl}, are discussed in detail in Ref.~\cite{Cirigliano:2024msg}. 
In this formalism, 
hard photon modes, with $E_\gamma \sim |\vec q_{\gamma} | \sim \Lambda_\chi$, where $\Lambda_\chi \sim 1$ GeV denotes the scale of chiral symmetry breaking, are encoded in the single nucleon $\gamma W$ box~\cite{Seng:2018qru}, and, at next-to-leading order, in two two-nucleon low-energy constants (LECs)~\cite{Cirigliano:2024msg}. 
Photons with momentum 
$|\vec q_{\gamma} | \sim
k_F$, where $k_F \sim 100$ MeV denotes the Fermi momentum, are sensitive to the structure of nuclei, and their contribution is encoded in the nuclear structure dependent
and Coulomb corrections, $\delta_{\rm NS}$  and $\delta_C$. These contributions can be 
expressed in terms of a well-defined set of nuclear matrix elements, 
which can be
calculated with nuclear many body methods, such as the nuclear shell model~\cite{Towner:1992xm}
or, more recently, {\it ab initio} many body methods based on chiral EFT~\cite{Gennari:2024sbn,Cirigliano:2024msg,King:2025fph}. 
Finally, photons with energy and momentum of order of the lepton energies, a few MeVs, can no longer resolve the nuclear structure, and they give rise to corrections which can be computed perturbatively in an expansion in $\alpha$
and  $E_e/k_F$. The most important correction in this class is given by the Fermi function~\cite{Fermi:1934hr},
which describes the wavefunction of the emitted electron or positron in the field of the final state nucleus. The Fermi function contains terms of $\mathcal O((\alpha Z)^n)$, where $Z$ is the charge of the final state nucleus. The work of Ref.~\cite{Hill:2023acw,Hill:2023bfh,Borah:2024ghn} clarified that the dependence of the Fermi function on a phenomenological nuclear radius, $R \sim k_F^{-1}$, can be reinterpreted as dependence on a renormalization scale $\mu$ which separates the photon modes sensitive to nuclear structure (soft and potential, in the language of nonrelativistic EFTs) from the low-energy ultrasoft modes. Ref.~\cite{Cirigliano:2024msg} then derived a definition for the electroweak operators whose matrix elements determine the exact form of the logarithm at the matching scale $\mu \sim k_F$.   The second important ultrasoft correction is given by the Sirlin function~\cite{Sirlin:1967zz}, which determines radiative corrections at $\mathcal O(\alpha)$.

The analysis of Refs.~\cite{Hill:2023acw,Cirigliano:2024msg,Borah:2024ghn}
pointed out that the large logarithms 
observed in the 
$\mathcal O(\alpha^2 Z)$ corrections also arise from the separation between the nuclear physics scale and the scale of the $\mathcal Q$ value, and can be predicted and resummed with renormalization group techniques. The logarithms match the corrections identified by W.~Jaus and G.~Rasche in Ref.~\cite{Jaus:1970tah} and then refined in Refs.~\cite{Sirlin:1986cc,Jaus:1986te}. The non-logarithmic terms in the $\mathcal O(\alpha^2 Z)$ corrections were estimated in the extreme relativistic and non-relativistic approximations in Ref.~\cite{Sirlin:1986cc}. However, the calculation of these terms in the EFT framework, and for generic electron and positron velocity $\beta \neq  0, 1$ is still missing. Ref.~\cite{Cirigliano:2024msg}  estimated the size of these corrections by varying the ultrasoft renormalization scale between $E_0$ and $4 E_0$, where $E_0 = \mathcal Q + m_e$ is the maximal energy of the electron/positron.
For $^{10}$C and $^{14}$O, the theoretical uncertainty induced by the ultrasoft scale variation was found to be about  $\left. \delta V_{ud} \right|_\mu \approx 2\cdot 10^{-4}$~\cite{Cirigliano:2024msg,King:2025fph}, subdominant with respect to the nuclear theory uncertainties, but still significant. 
Since it scales with $Z$,  for heavier system such as $^{46}$V or $^{54}$Co, this uncertainty becomes as big as the one induced by $\delta_{\rm NS}$. 
To remediate this situation, in this paper we calculate $\mathcal O(\alpha^2 Z)$ corrections to superallowed $\beta$ decays, in a low-energy heavy-particle EFT
with nuclear degrees of freedom.

The paper is organized as follows. In Section~\ref{sec:Lagrangian}
we introduce the heavy-particle Lagrangian relevant for the ultrasoft corrections to $0^+\to 0^+$ transitions, provide details on the matching coefficient $C_{\text{eff}}^{(g_V)}\left(\mu\right)$ and review  $\mathcal{O}(\alpha)$ results. 
In Section~\ref{sec:results} we summarize the main results of this paper, namely the $\mathcal O(\alpha^2 Z)$ corrections to the decay rate and to the electron-neutrino asymmetry.
In Section~\ref{sec:pheno} we discuss the phenomenological implications of our results. In Section~\ref{sec:computational} we provide more details on the computational framework used to derive the results presented in Section~\ref{sec:results}, and we conclude in Section~\ref{sec:conclusion}. In Appendix~\ref{sec:countertermlagrangian} we discuss the renormalization of the theory, in Appendix~\ref{sec:masterintegrals} we provide the differential equations and analytic results for the master integrals, and finally in Appendix~\ref{app:alpha2} we present the differential equations for the master integrals contributing at $\mathcal{O}(\alpha^2).$

\section{Heavy particle Lagrangian for $0^+ \rightarrow 0^+$ transitions }\label{sec:Lagrangian}

Ultrasoft photons cannot resolve the individual nucleons inside a nucleus and see the nucleus as a whole, thus becoming sensitive to the global nuclear properties like the charge or the charge radius.
The contribution of ultrasoft photons can thus be computed in a low-energy theory with nuclear degrees of freedom, minimally coupled  to the photon. 
Since the nuclear mass never plays a dynamical role, 
it is further convenient to work in a heavy particle formalism~\cite{Georgi:1990um,Jenkins:1990jv}. At leading order in $1/k_F$, the  Lagrangian reduces to 
\begin{align}\label{eq:Lag1}
    \mathcal L &= \mathcal L_{\text{QED}} + \mathcal L_{e^-} + \mathcal L_{e^+}, 
\end{align}
where $\mathcal L_{\text{QED}}$ is the QED Lagrangian
\begin{align}
    \mathcal{L}_{\text{QED}}=- \frac{1}{4}   F^{\mu \nu} F_{\mu \nu} + \bar \nu i \slashed{\partial} \nu +
    \bar e  \left( i \slashed{D} - m_e\right)  e,
\end{align}
while $\mathcal L_{e^-}$ and $\mathcal L_{e^+}$ are the Lagrangians for electron and positron emitters, respectively (see also Ref.~\cite{Plestid:2024eib})
\begin{align}
    \mathcal L_{e^-} &=   \mathcal{A}_f^\dagger \, \left( i v\cdot D + \Delta \right) {\mathcal A}_f + \mathcal A_i^\dagger\, i v\cdot D {\mathcal A}_i   - \frac{2 G_F}{\sqrt{2}} V_{ud} \left(  C_V \mathcal A_f^\dagger v_\mu {\mathcal A}_i \, \bar e \gamma^\mu P_L \nu + {\rm h.c.}\right), \nonumber \\
  \mathcal L_{e^+} &=  
\mathcal{B}_f^\dagger\, \left( i v\cdot D + \Delta \right) {\mathcal B}_f + \mathcal B_i^\dagger\, i v\cdot D {\mathcal B}_i   - \frac{2 G_F}{\sqrt{2}} V_{ud} \left(  C_V \mathcal B_f^\dagger v_\mu {\mathcal B}_i \, \bar \nu \gamma^\mu P_L e + {\rm h.c.}\right). \label{eq:Lpm}
\end{align}
We have introduced here two complex scalar fields 
$\mathcal A_{i, f}$ ($\mathcal B_{i, f}$) that describe  the initial and final state $0^+$ nuclei.  
The $0^+$ nuclei relevant to the extraction of $V_{ud}$ belong to isospin $1$ multiplets. This information could be incorporated in  Eq.~\eqref{eq:Lpm} by adding isospin indices to the fields. In the low-energy EFT, however, isospin is broken both by the mass splitting $\Delta$ 
and by the coefficient $C_V$, which, as we will discuss, is nucleus-dependent. We thus find convenient to add a separate pair of scalar fields for each superallowed transition.
In Eq.~\eqref{eq:Lpm}, $v^{\mu} = (1, \vec{0})$ is the nuclear velocity vector and 
$D_\mu$ is the covariant derivative
\begin{equation}
    D_\mu = \partial_\mu + i e Q A_\mu.
\end{equation}
For both electron and positron emitters,  we denote by $Q_f = Z$ the charge of the final state nucleus. $Q_i = Z\mp 1$ is the charge of the initial state nucleus, where the upper sign is for electron decays and the lower for positron decays. For the electron field, $Q_e = -1$.
In the heavy particle formalism, we have some freedom in removing the inert mass scale. Here we choose to measure the mass with respect to the mass of the initial state. $\Delta$ thus denotes the mass difference between initial and final state,
\begin{equation}
    \Delta = m_n - m_p + B_f - B_i = \mathcal Q + m_e,
\end{equation}
where the binding is defined to be positive, and $\mathcal Q$ is the reaction's $\mathcal Q$ value.
The on-shell relation is then  $v \cdot p = 0$ for the initial state and $v \cdot p^\prime = -\Delta$ for the final state nuclei.
The last term in Eq.~\eqref{eq:Lpm} mediates the weak decay. $C_V$ denotes the Fermi matrix elements.
At leading order in $\alpha$ we have
$C_V = M_{F}^{(0)} = \sqrt{2}$. At higher orders, $C_V$ receives corrections from integrating out hard, soft and potential photon modes~\cite{Cirigliano:2024msg},
and acquires a dependence on the renormalization scale. 
Ref.~\cite{Cirigliano:2024msg} worked in the $\overline{\text{MS}}_\chi$ scheme used in the chiral perturbation theory literature~\cite{Gasser:1983yg}, 
which can be related to $\overline{\text{MS}}$ by defining the scale
\begin{equation}
    \bar\mu = \bar\mu_\chi e^{-1},
\end{equation}
where $\bar\mu$ is the renormalization scale in $\overline{\text{MS}}$.
Ref.~\cite{Cirigliano:2024msg} chose a matching scale
\begin{equation}\label{eq:matchscale}
    \mu_\pi = R^{-1} \exp \left(\frac{1}{2} - \gamma_E \right),
\end{equation}
with $
    R^2 = \frac{5}{3} \langle r^2 \rangle
$
and $\langle r^2 \rangle$ denotes the nuclear charge radius.
At the matching scale, the vector coupling can be expressed as 
\begin{equation}
    C_V(\bar\mu_\chi = \mu_\pi) = M_F^{(0)} C_{\text{eff}}^{(g_V)}(\mu) \left(1 - \frac{1}{2} \delta_C + \frac{1}{2} \delta^{(0)}_{\rm NS}\right).   
\end{equation}
In this expression, $\delta_C$ and $\delta_{\rm NS}$
denote two corrections that depend on the nuclear structure. They are different for different nuclei and need to be calculated using nuclear many body methods. As our focus is on the perturbatively calculable corrections, we will set $\delta_C = \delta_{\text{NS}} = 0$ in what follows.
The matching coefficient  $C^{(g_V)}_\text{eff}$ contains information on the single nucleon $\gamma W$ box, and on additional contributions arising at the scale $\mu_\pi$. Ref.~\cite{Cirigliano:2024msg} expressed $C_{\text{eff}}^{(g_V)}$ as
\begin{align}\label{eq:Ceff}
C_\text{eff}^{(g_V)}(\mu_\pi) &= g_V(\mu_\pi)\left[c_W^{(g_V,0)}+Z c_W^{(g_V,1)}+ Z^2 c_W^{(g_V,2)}\right].
\end{align}
Here $g_V$ is the single nucleon vector coupling. 
Using nonperturbative input on the $\gamma W$ box from Refs.~\cite{Seng:2018qru,Seng:2018yzq,Czarnecki:2019mwq,Shiells:2020fqp,Hayen:2020cxh,Seng:2020wjq,Cirigliano:2022yyo}, Ref.~\cite{Cirigliano:2023fnz} found
\begin{equation}
    \label{eq:gVmu0}
g_V(\bar\mu_\chi = M_{\pi^\pm})= 1.01494(12), 
\end{equation}
where $M_{\pi^\pm}$ is the mass of the charged pions and the uncertainty is dominated by the nonperturbative component of the $\gamma W$ box. Perturbative $\mathcal O(\alpha \alpha_s)$  corrections to $g_V$ were recently obtained in Ref.~\cite{Moretti:2025qxt}. We do not include them in this analysis, but it will be important to consider them in extractions of $V_{ud}$. 

The remaining terms in Eq.~\eqref{eq:Ceff} arise from integrating out potential photons in chiral EFT. They are given by
\begin{align}
c_W^{(g_V,2)}=C_\delta^\text{3b}, \quad
 c_W^{(g_V,1)}= C_\delta - 2 C_\delta^\text{3b}  \mp C_\delta^\text{3b},  \quad 
  c_W^{(g_V,0)}=1+(-1\mp 1)\left(\frac{1}{2}C_\delta -C_\delta^\text{3b}\right),
\end{align}
where the top (bottom) sign is for electron (positron) decay.
The two matching coefficients $C_\delta$ and $C_{\delta}^{\text{3b}}$ were obtained in Ref.~\cite{Cirigliano:2024msg} by calculating the electroweak potential at $\mathcal O(\alpha^2)$. They are given by
\begin{align}
        C_\delta &=  -\frac{\alpha^2}{2} \left(  \log \frac{\bar\mu^2_\chi}{\Lambda^2} -\frac{13}{8}+ 2\gamma_E \right), \label{eq:cdelta}  \\
    C_{\delta}^\text{3b} &= -\alpha^2 \left( \frac{1}{4}\log \frac{\bar\mu^2_\chi}{ \Lambda^2} + \frac{\gamma_E}{2}  - \frac{3}{8}\right),
    \label{eq:cdelta3b}
\end{align}
where $\Lambda$ is an arbitrary cut-off. The dependence on $\Lambda$ cancels between $C_{\text{eff}}^{(g_V)}$ and $\delta_{\text{NS}}$~\cite{Cirigliano:2024msg}.

The coefficient $C^{(g_V)}_{\text{eff}}$ satisfies the renormalization group equation~\cite{Borah:2024ghn}
\begin{align}\label{eq:CeffRG}
\frac{d C^{(g_V)}_\text{eff}(\mu)}{d\log \mu} &= \gamma^{(g_V)}C_\text{eff}^{(g_V)}(\mu),\\
\gamma^{(g_V)}&=\frac{\alpha}{\pi}\tilde \gamma_0+\left(\frac{\alpha}{\pi}\right)^2\tilde \gamma_1  +\left[\sqrt{1-\alpha^2 Z(Z\mp1)}-1\right] + \frac{\alpha^3}{4\pi}Z^2\left(6-\frac{\pi^2}{3}\right),
\end{align}
for $\beta^\mp$ decays. The anomalous dimensions $\tilde \gamma_{0,1}$ are given by
\begin{equation} \label{eq:anomalous}
\tilde \gamma_0 = -\frac{3}{4},\quad \tilde \gamma_1 = \frac{5 \tilde{n}}{24} + \frac{5}{32} - \frac{\pi^2}{6},
\end{equation}
where $\tilde  n = 1$ for scales $\mu \le M_{\pi^\pm}$.

Eq.~\eqref{eq:Lag1} is the first term in an expansion in $E_e/k_F$. 
The next correction arises at $\mathcal O( \alpha Z E_e R)$, and, in the formalism of Refs.~\cite{Hardy:2020qwl,Hayen:2017pwg}, is captured by the finite size corrections $L_0(Z,E_e)$ and by the shape 
factor $C(Z,E_e)$ (see Ref.~\cite{Hayen:2017pwg}). In the approach of Ref.~\cite{Cirigliano:2024msg}, these corrections were encoded by the energy-dependent electroweak operators $\mathcal V^0_E$ and $\mathcal V_{m_e}$, which can also be matched onto a low-energy effective Lagrangian. For example, for electron emitter, one would have
\begin{equation}
    \mathcal L^{(1)} = -\frac{2 G_F}{\sqrt{2}}  V_{ud} 
   {\mathcal A}^\dagger_f \mathcal{A}_i \left[ 
C^{}_{V E_0} \, i v \cdot D \left(
\bar e_L \slashed{v} P_L \nu_L\right ) 
+ C^{}_{V E_e}   \bar e_L v\cdot \overleftarrow{D} \slashed{v} P_L \nu_L
+ m_e C_{V m_e} \bar e_R  P_L \nu_L
 \right].
\end{equation}
The coefficients $C_{V E_0}$, $C_{V E_e}$ and $C_{V m_e}$
have dimension of $[\text{mass}]^{-1}$ 
and are determined by  the calculation of nuclear matrix elements. As can be seen in Refs.~\cite{Cirigliano:2024msg,King:2025fph}, these matrix elements  
scale as $\alpha Z R$. At order $R^2$ additional contributions from the nuclei electromagnetic and weak radii appear.
The methods presented here can be  extended to calculate radiative corrections arising from subleading terms in the heavy particle Lagrangian, which need to be included for determinations of $V_{ud}$ at sub-permille accuracy.

\subsection{Tree level decay rate}

Using the Lagrangian~\eqref{eq:Lag1}, the tree level decay rate is given by 
\begin{align}
    \frac{d \Gamma}{d E_e d\cos\theta_{e\nu}} =
\frac{1}{4\pi^3} G_F^2 V^2_{ud} C_V^2  p_e E_e  (E_0 - E_e)^2  (1 + \beta \cos\theta_{e\nu}),
\end{align}
where $E_0 = \mathcal Q + m_e$ is the electron or positron endpoint energy, $p_e = |\vec p_e|$, 
$\cos\theta_{e\nu}$ is the angle between the charged lepton  and neutrino momenta 
\begin{equation}
    \cos\theta_{e\nu} = \frac{\vec p_e \cdot \vec p_\nu}{|\vec p_e|\,  E_\nu},
\end{equation}
and $\beta$ is the charged lepton velocity
\begin{equation}
    \beta = \frac{|\vec p_e|}{E_e} = \sqrt{1 - \frac{m_e^2}{E_e^2}}.
\end{equation}
Once we include electromagnetic radiative corrections, we will write\footnote{As the neutrino momentum $\vec p_\nu$ and the angle $\theta_{e\nu}$ are not directly observable, the function  $a(\beta,\bar E)$ represents corrections to an idealized electron-neutrino asymmetry. To be phenomenologically relevant, these corrections need to be adapted  either to the recoil or to pseudo-neutrino formalism, see the discussion in Ref.~\cite{Seng:2023ynd}.
We thank C.~Y.~Seng for stressing this point. 
}
\begin{align}
    \frac{d \Gamma}{d E_e d\cos\theta_{e\nu}} &=
\frac{1}{4\pi^3} G_F^2 V^2_{ud} C_V^2 \, p_e E_e \bar E^2  \Bigg[ (1 + \beta\cos\theta_{e\nu}) F(\beta)  +  g(\beta,\bar E)  + a(\beta,\bar E) \beta \cos\theta_{e\nu}  \Bigg],
\end{align}
where $\bar E = E_0 - E_e$. $F(\beta)$ captures corrections subsumed by the Fermi function~\cite{Fermi:1934hr}, and has the expansion
\begin{align}
    F(\beta,\bar\mu) = \sum_{n=0}^\infty (\alpha Z)^n f^{(n)}(\beta,\bar\mu),
\end{align}
while the functions $g$ and $a$ capture subleading corrections in $\alpha$ and $Z$.
\begin{align}
    g(\beta, \bar E,\bar \mu) &= \sum_{n=1}^\infty \alpha^n Z^{n-1} g^{(n)}(\beta, \bar E, \bar\mu) + \sum_{n=2}^\infty \alpha^n Z^{n-2} h^{(n)}(\beta,\bar E,\bar\mu) + \ldots \\
    a(\beta, \bar E,\bar\mu) &= \sum_{n=1}^\infty \alpha^n Z^{n-1} a^{(n)}(\beta, \bar E,\bar \mu) + \sum_{n=2}^\infty \alpha^n Z^{n-2} b^{(n)}(\beta,\bar E,\bar\mu) + \ldots.
\end{align}
Here $\alpha\equiv \alpha(\bar\mu)$ denotes the running electromagnetic coupling in the $\overline{\text{MS}}$ scheme.
The functions $F(\beta,\bar\mu)$, $g(\beta, \bar E,\bar\mu)$ and $a(\beta,\bar E,\bar\mu)$  depend on the renormalization scale in such a way as to compensate for the scale dependence of $C_V$.
The goal of this paper is to compute the functions $g^{(2)}$ and $a^{(2)}$.
As the perturbative expansion of the Fermi function does not converge very well for the values of $Z$ relevant for the extraction of $V_{ud}$, it is customary in the literature to include it at all orders and to factor it out from other electromagnetic corrections. We will then also provide the expressions
\begin{align}\label{eq:factorized}
    \frac{d \Gamma}{d E_e d\cos\theta_{e\nu}} &=
\frac{1}{4\pi^3} G_F^2 V^2_{ud} C_V^2 \,  p_e E_e \bar E^2 F(\beta)  \Bigg[  1 + \alpha\, \hat{g}^{(1)}(\beta,\bar E) + \alpha^2 Z\, \hat{g}^{(2)}(\beta,\bar E) \nonumber \\ & 
+  \beta \cos\theta_{e\nu}  \left( 1 + \alpha\, \hat{a}^{(1)}(\beta,\bar E) +
\alpha^2 Z \, \hat{a}^{(2)}(\beta,\bar E)
\right) \Bigg],
\end{align}
with $\hat{g}^{(1)} = g^{(1)}$, $\hat{a}^{(1)} = a^{(1)}$ and
\begin{align}
    \hat{g}^{(2)}(\beta,\bar E,\bar\mu) &= g^{(2)}(\beta, \bar E,\bar\mu) - f^{(1)}(\beta,\bar\mu) \, g^{(1)}(\beta, \bar E,\bar\mu), \\
 \hat a^{(2)}(\beta,\bar E,\bar\mu) &=   a^{(2)}(\beta, \bar E,\bar\mu) - f^{(1)}(\beta,\bar\mu) \, a^{(1)}(\beta, \bar E,\bar\mu) .
\end{align}

\subsection{$\mathcal O(\alpha Z)$ and $\mathcal O(\alpha)$ corrections}
The one-loop results are well known~\cite{Wilkinson:1982hu,Sirlin:1967zza}.
The leading $\alpha Z$ dependence is given by 
\begin{align}\label{eq:f1}
    f^{(1)}(\beta) &= \pm \frac{\pi}{\beta},
\end{align}    
where the $\pm$ sign correspond to electron and positron emissions, respectively. 
This is the first term of the expansion of the  Fermi function in powers of $\alpha$.
The calculation of the Fermi function in an EFT formalism was developed in Refs.~\cite{Hill:2023acw,Hill:2023bfh}. Here the Fermi function was calculated at all orders in $\alpha Z$, giving
 \begin{align}\label{eq:FF}
\bar{F}(\beta,\bar\mu)  &= \frac{4\eta}{(1+\eta)^2}\frac{2(1+\eta)}{\Gamma(2\eta+1)^2}|\Gamma(\eta+iy)|^2 e^{\pi y}\times \left(\frac{2 |\vec p_e|}{\bar\mu}e^{-\gamma_{E}}\right)^{2(\eta-1)},
\end{align}
with $\eta = \sqrt{1-\alpha^2 Z^2}$ and $y =\pm Z\alpha/\beta$ 
for $\beta^{\mp}$ decays.
This expression can be identified with the traditional Fermi function by setting the $\overline{\rm MS}$ renormalization scale to $\bar\mu = R^{-1} e^{-\gamma_{E}}$, where $R$ is the nuclear radius. After this identification, Eq.~\eqref{eq:FF} differs from the traditionally employed Fermi function by $4\eta/(1+\eta)^2\approx1-\alpha^4Z^4/16$. 
The first corrections to the $g$ and $a$ functions are given by~\cite{Sirlin:1967zza,Ando:2004rk}
\begin{align}
    g^{(1)}(\beta,\bar E) & =  \frac{1}{2\pi} \Bigg\{ \frac{3}{2} L_\mu  - \frac{4}{\beta}\left({\rm Li}_2\left(\frac{2\beta}{1+\beta}\right) + \frac{1}{4}\log^2 \frac{1+\beta}{1-\beta} \right) 
    + 2 \log \frac{m_e^2}{4\bar E^2}
+ 8 - \frac{4}{3} \frac{\bar E}{E_e}    
    \nonumber \\ + 
& \frac{1}{\beta} \log \frac{1+\beta}{1-\beta}  \left( - \log \frac{m_e^2}{4 \bar E^2} -2 + \beta^2 + \frac{\bar E^2}{12 E_e^2} + \frac{2}{3} \frac{\bar E}{E_e}\right)
    \Bigg\}, \label{eq:Sirlin1} \\
    a^{(1)}(\beta,\bar E)&= g^{(1)}(\beta,\bar E) +  \frac{1}{2\pi} \Bigg\{ \frac{1-\beta^2}{\beta^2}  \frac{1}{\beta} \log \frac{1+\beta}{1-\beta}  \left(\frac{\bar E^2}{12 E_e^2} + \frac{2}{3} \frac{\bar E}{E_e} + \beta^2\right) \nonumber \\
    &  \frac{4}{3} \frac{\bar E}{E_e} - \frac{\bar E^2}{6 E_e^2 \beta^2} - \frac{4}{3} \frac{\bar E}{\beta^2 E_e}\label{eq:Sirlin2}
    \Bigg\}.
\end{align}
Here $L_\mu = \log \bar\mu^2/m_e^2$. Eq.~\eqref{eq:Sirlin1} agrees with the Sirlin function~\cite{Sirlin:1967zza}, as computed in Heavy Baryon EFT~\cite{Cirigliano:2022hob,Ando:2004rk}. 
As pointed out in Ref.~\cite{Cirigliano:2022hob}, this differs  from the Sirlin function computed in Ref.~\cite{Sirlin:1967zza} by a constant term, $11/4$. 
Similarly, the corrections to the electron-neutrino correlation agree with Refs.~\cite{Cirigliano:2022hob,Ando:2004rk}.

\section{Results}\label{sec:results}

We summarize here the results for the $\mathcal O(\alpha^2 Z^2)$ and $\mathcal O(\alpha^2 Z)$ terms in the virtual-virtual and real-virtual diagrams shown in Figs.~\ref{fig:diagsvirtualvirtual}
and~\ref{fig:virtualreal}. 
$\mathcal O(\alpha^2 Z^2)$ contributions 
are only induced by the virtual-virtual diagrams.
Combining with the square of the one-loop diagram, we obtain
\begin{equation}\label{eq:fermi2}
    f^{(2)}(\beta) = \left( \frac{\pi^2}{3\beta^2} + \frac{11}{4} + L_\beta \right) ,
\end{equation}
with
$L_\beta = \log \bar\mu/2 E_e \beta$.
This expression reproduces the $\mathcal O(\alpha^2)$ expansion of Eq.~\eqref{eq:FF}.

The $\mathcal O(\alpha^2 Z)$ corrections to $g$ and $a$ are the main result of our work.
The corrections to the $g$ function are given by
\begin{align}
&    g^{(2)}(\beta,\bar E) = \pm \Bigg\{
\left( \frac{5}{12 \beta} - \frac{1}{2}  \right) L_\mu - \frac{1}{2\beta} \left(2  - \frac{1}{\beta} \log\frac{1+\beta}{1-\beta}\right) \log \frac{4\bar E^2}{m_e^2}  \nonumber \\ & + \frac{4+9\beta -\beta^3}{2\beta^2} \PolyLog\left(
    \frac{1-\beta}{1+\beta}\right)  +\frac{-5 + \beta^2}{\beta} \PolyLog\left(\sqrt{  \frac{1-\beta}{1+\beta}}\right) - \frac{1}{\beta} \PolyLog\left( \left( \frac{1-\beta}{1+\beta}\right)^2\right) 
    \nonumber \\
&
- \frac{4-3\beta+ \beta^3}{2\beta^2}  \frac{\pi^2}{6} 
   - \frac{12-5\beta + \beta^3}{16\beta^2} \log^2 \frac{1+\beta}{1-\beta}  - \frac{2}{\beta^2} \log \frac{1+\beta}{1-\beta} \log \frac{2\beta}{1+\beta}  - \frac{2\beta^2 + 2}{\beta^2} \log \frac{1+\beta}{2}
   \nonumber \\
&
   + \frac{\beta^2-2}{\beta^2} \log \left( 1 + \sqrt{\frac{1-\beta}{1+\beta}}\right)    + \frac{1}{12 \beta^2} \log \frac{1+\beta}{1-\beta} \left( -18 + 10 \beta + 9 \beta^2 + 3 \beta^3 + \frac{\bar E^2}{2 E_e^2}  + 4 \frac{\bar E}{E_e}\right) \nonumber \\ &  - \frac{2}{3\beta} \frac{\bar E}{E_e}  + \left(1-\sqrt{\frac{1-\beta}{1+\beta}}\right)^3 \frac{(1+\beta)^2}{144\beta^4} \Bigg(  \sqrt{\frac{1-\beta}{1+\beta}}  (430-220 \beta-39\beta^2+48\beta^3)  \nonumber \\ &  + 434 - 652 \beta + 327 \beta^2 - 96 \beta^3 \Bigg) \Bigg\},
\label{eq:g2}
\end{align}
for electron and positron emissions, respectively.
Factoring out the Fermi function as in Eq.~\eqref{eq:factorized}, we find
\begin{align}
&       \hat{g}^{(2)}(\beta,\bar\mu) = \pm \Bigg\{
-\left( \frac{1}{3 \beta} + \frac{1}{2}  \right) L_\mu   + \frac{9 -\beta^2}{2\beta} \PolyLog\left(
    \frac{1-\beta}{1+\beta}\right) +  \frac{-5 + \beta^2}{\beta} \PolyLog\left(\sqrt{  \frac{1-\beta}{1+\beta}}\right) \nonumber \\
& - \frac{1}{\beta} \PolyLog\left( \left( \frac{1-\beta}{1+\beta}\right)^2\right) 
+ \frac{3 - \beta^2}{2\beta}  \frac{\pi^2}{6}
    - \frac{4-5\beta + \beta^3}{16\beta^2} \log^2 \frac{1+\beta}{1-\beta}   + \frac{\beta^2-2}{\beta^2} \log \left( 1 + \sqrt{\frac{1-\beta}{1+\beta}}\right) \nonumber \\
& - \frac{2\beta^2 + 2}{\beta^2} \log \frac{1+\beta}{2}   + \frac{1}{12 \beta^2} \log \frac{1+\beta}{1-\beta} \left( -6 + 10 \beta + 3 \beta^2 + 3 \beta^3 \right) - \frac{4}{\beta}  \nonumber  \\
& + \left(1-\sqrt{\frac{1-\beta}{1+\beta}}\right)^3 \frac{(1+\beta)^2}{144\beta^4} \Bigg(  \sqrt{\frac{1-\beta}{1+\beta}}  (430-220 \beta-39\beta^2+48\beta^3) \nonumber \\ &   + 434 - 652 \beta + 327 \beta^2 - 96 \beta^3 \Bigg) \Bigg\},
\label{eq:dg2}
\end{align}
where again the upper sign is for electrons, the lower for positrons.
Notice that all the dependence on $\bar E$, which arises from the real emission diagrams, drops out in $\hat g^{(2)}$. At $\mathcal O(\alpha^2 Z)$, the dependence on $\bar E$ is thus fully contained in the product of the one-loop Fermi and Sirlin functions. 
The dependence on the renormalization scale in Eqs.~\eqref{eq:g2} and \eqref{eq:dg2} arises from two sources.
The logarithm proportional to $1/\beta$ compensates the scale dependence of  the couplings in the leading $\mathcal O(\alpha Z)$ term in Eq.~\eqref{eq:f1},
which is due to the $\overline{\text{MS}}$ electromagnetic coupling $\alpha$ and to the one-loop running of $C_V$.
The latter cancel in $\hat{g}_2$, so that the logarithm proportional to $1/\beta$ is solely determined by the running of $\alpha$. The remaining logarithm is determined by the two-loop running
of $C_V$.

The correction to the electron-neutrino asymmetry depends on the definition of the electron mass. Here $m_e$ denotes the electron pole mass, and, with this definition, we get
\begin{align}
 &   a^{(2)}(\beta,\bar E,\bar\mu) = g^{(2)}(\beta,\bar E,\bar\mu)   \pm \Biggl\{\frac{(1-\beta^2)^2}{\beta^3} \Bigg( \PolyLog{\left(-\sqrt{\frac{1-\beta}{1+\beta}}\right) } + \frac{1}{16} \log^2 \frac{1+\beta}{1-\beta} + \frac{\pi^2}{12} \Bigg)  \nonumber \\ & - 3 \frac{1-\beta^2}{\beta^2} \log\left( 1 + \sqrt{\frac{1-\beta}{1+\beta}} \right)  - \frac{2 (1-\beta^2)}{\beta^2} \log\frac{1+\beta}{2}   - \frac{\bar E^2}{12\beta^3 E_e^2}  - \frac{2}{3} \frac{1-\beta^2}{\beta^3} \frac{\bar E}{E_e} \nonumber \\
    & + \frac{1-\beta^2}{\beta^4} \log \frac{1+\beta}{1-\beta} \left( \frac{\bar E^2}{24 E_e^2} + \frac{\bar E}{3 E_e} + \beta \frac{2 - \beta+  \beta^2}{4}\right) +  (1-\beta^2)^{\frac{3}{2}} \frac{1+2\beta^2}{3\beta^4} \nonumber \\ & + \frac{1-\beta^2}{6\beta^4} (-2 - 9 \beta^2 + 12 \beta^3)\Biggr\}, 
\end{align}
and
\begin{align}
  &  \hat{a}^{(2)}(\beta, \bar \mu)  = \hat{g}^{(2)}(\beta,\bar\mu) 
  \pm \frac{1-\beta^2}{\beta^2} \Bigg\{
        \frac{(1-\beta^2)}{\beta} \Bigg( \PolyLog{\left(-\sqrt{\frac{1-\beta}{1+\beta}}\right) } + \frac{1}{16} \log^2 \frac{1+\beta}{1-\beta} + \frac{\pi^2}{12} \Bigg) \nonumber \\ &    - 3  \log\left( 1 + \sqrt{\frac{1-\beta}{1+\beta}} \right)  - 2 \log\frac{1+\beta}{2}      + \frac{2-3\beta + \beta^2}{4} \frac{1}{\beta} \log\frac{1+\beta}{1-\beta}+  (1-\beta^2)^{\frac{1}{2}} \frac{1+2\beta^2}{3\beta^2} \nonumber \\ &
+ \frac{1}{6\beta^2} (-2 - 9 \beta^2 + 12 \beta^3) \Bigg\}. \label{eq:da2}
\end{align}
Also $\hat{a}^{(2)}$ is independent of $\bar E$. 
Notice that in the massless limit, $\beta =1$, $\hat{a}^{(2)} = \hat{g}^{(2)}$.

\subsection{Comparison with the literature}

We can compare Eq.~\eqref{eq:dg2} with the $\mathcal O(\alpha^2 Z)$ corrections obtained in Ref.~\cite{Sirlin:1986cc}, which presented results in the relativistic and non-relativistic limits, corresponding, respectively, to $\beta\rightarrow 1$ and $\beta \rightarrow 0$. The authors of Ref.~\cite{Sirlin:1986cc}
used relativistic propagators for the nucleon undergoing $\beta$ decay, and then expanded the results in powers of $1/m_N$. 
By introducing a nuclear form factor, the logarithm of the nucleon mass is subsequently replaced by logarithms of the nuclear radius.
To compare with Ref.~\cite{Sirlin:1986cc}, we evaluate the correction $\hat{g}^{(2)}$ in the relativistic and non-relativistic limit, obtaining
\begin{align}
    \delta g^{(2)}(\beta,\bar E) &\xrightarrow[\beta \rightarrow 0]{} \pm\left[-\left( \frac{1}{3\beta}  + \frac{1}{2}\right) L_\mu - \frac{37}{8} + 4 \log(2)\right],  \label{eq:nonrelativistic} \\
    \delta g^{(2)}(\beta,\bar E) &\xrightarrow[\beta \rightarrow 1]{} \pm \left[- \frac{5}{6} \log \frac{\mu^2}{4 E_e^2} - \frac{131}{36} + \frac{\pi^2}{6} \right], \label{eq:relativistic}
 \end{align}
for electron and positron emitters. 
We can further break up these results in a component coming
from the vacuum polarization and vertex correction diagrams
(and the respective counterterms)  (diagrams $(g)$ to $(l)$ in Fig.~\ref{fig:diagsvirtualvirtual})
and from the remaining real and virtual diagrams in Figs.~\ref{fig:diagsvirtualvirtual} and~\ref{fig:virtualreal}.
In the relativistic limit, we obtain
\begin{align}
    \left. \delta g^{(2)}(\beta,\bar E) \right|_{(g)- (l)}&\xrightarrow[\beta \rightarrow 1]{} \pm \left[ \frac{1}{9} - \frac{\pi^2}{6} - \frac{1}{3} \log \frac{\mu^2}{4 E_e^2} \right], \label{eq:compvac} \\
    \left. \delta g^{(2)}(\beta,\bar E) \right|_{\text{other}}&\xrightarrow[\beta \rightarrow 1]{} \pm \left[ -\frac{15}{4} + \frac{\pi^2}{3} - \frac{1}{2} \log \frac{\mu^2}{4 E_e^2} \right]. \label{eq:compother}
\end{align}
As the logarithm in Eq.~\eqref{eq:compvac}
is associated with the running of $\alpha$, its natural scale is $\mu \sim m_e$. The logarithm in Eq.~\eqref{eq:compother}, on the other hand, compensates the running of $C^{(g_V)}_{\text{eff}}$, and thus its natural scale is, as we will see shortly, the nuclear scale $R^{-1}$.
With these identifications, one can see that the logarithms in Eqs.~\eqref{eq:compvac} and~\eqref{eq:compother} match the expressions in Ref.~\cite{Sirlin:1986cc}.  In addition, the finite parts of Eq.~\eqref{eq:compvac} agree with the calculation of Ref.~\cite{Sirlin:1986cc}. 
On the other hand, we find that Eq.~\eqref{eq:compother} differs from the corresponding result in Ref.~\cite{Sirlin:1986cc}, which, for positron decays, obtained
\begin{align}
     \left. \delta g^{(2)}(\beta,\bar E) \right|_{\text{other}}& \xrightarrow[\beta \rightarrow 1]{}   \frac{5}{2} - \frac{\pi^2}{6}.
\end{align}
We notice that, while diagrams $(g)$ to $(l)$  
are finite after charge and mass renormalization, in the heavy particle formalism  
diagrams $(a)$ to $(f)$ are ultraviolet divergent, and require further renormalization. The difference between finite pieces might thus arise from the different ultraviolet regulator used in the two calculations, namely dimensional regularization versus a relativistic propagator. 
As already remarked, similar differences in the finite pieces appear in the Sirlin function at one loop, and Ref.~\cite{Sirlin:1986cc} also noticed that the logarithm of the nucleon mass, and the accompanying finite pieces, change when including nuclear finite size effects, which is equivalent to changing regularization.
The difference between our result and Ref.~\cite{Sirlin:1986cc} is thus likely to denote a scheme dependence. This can be absorbed by the definition of the coupling $C_V$, which needs to be computed in a consistent scheme.   
$\mathcal O(\alpha^2 Z)$ corrections to superallowed decays 
in the relativistic limit 
were very recently computed in Ref.~\cite{Plestid:toappear}. This work also uses a heavy-particle effective field theory formalism. 
Our result in Eq.~\eqref{eq:relativistic} agrees with Ref.~\cite{Plestid:toappear}, confirming that the discrepancy with Ref.~\cite{Sirlin:1986cc} is likely due to the ultraviolet regulator.

To explicitly manifest the origin of the logarithms in Eqs.~\eqref{eq:nonrelativistic} and~\eqref{eq:relativistic}, we can include the $\mathcal O(\alpha^2 Z)$ piece of $C_V$.
Focusing on positron emission, and turning off the resummation in the matching coefficient, we find
\begin{align}
    C^2_V  (1 + \delta g_2(\beta, \bar E)) &\xrightarrow[\beta \rightarrow 0]{}   \frac{1}{3\beta} \log \frac{\mu^2}{m_e^2}  - \frac{1}{2} \log \frac{m_e^2}{\Lambda^2}  + \frac{11}{2} - 4 \log(2) - \gamma_E, \label{eq:combined1} \\
    C^2_V  (1 + \delta g_2(\beta, \bar E)) &\xrightarrow[\beta \rightarrow 1]{}   \frac{1}{3} \log \frac{\mu^2}{4E_e^2}  - \frac{1}{2} \log \frac{4E_e^2}{\Lambda^2}  + \frac{325}{72} - \frac{\pi^2}{6}  - \gamma_E.\label{eq:combined2}
\end{align}
We see that the $\beta$-independent $\log(\mu)$ is replaced by a logarithm of the nuclear scale $\Lambda$. The expressions in Eqs.~\eqref{eq:combined1} and~\eqref{eq:combined2} make it clear that the finite pieces depend on the exact value of $\Lambda$, which in turn requires the calculation of the nuclear matrix element of the electroweak operator denoted as $\mathcal V_+$ in Ref.~\cite{Cirigliano:2024msg}.

\section{Phenomenological implications}
\label{sec:pheno}

\begin{table}[]
    \centering
    \begin{tabular}{||c|c|c||c||c|c||}
    \hline
       Transition  & $\mathcal{Q}_{EC}$\,(KeV) &  $\mu_\pi$\,(MeV) & $\overline{P}( \alpha)$   & $ \overline{P}(\alpha, \alpha^2 Z)$ \\
       \hline
$^{10}$C $\rightarrow$ $^{10}$B &  $1907.99(7)$ & 58.3 & $2.39118(57)_{g_V}(85)_{\mu}$  & $2.39280(57)_{g_V}(0)_{\mu}$\\
\hline 
$^{14}$O $\rightarrow$ $^{14}$N  & 
  $2831.54(8)$ & 55.3 & $44.375(11)_{g_V}(20)_\mu$ & $44.417(11)_{g_V}(0)_\mu$ \\
\hline
$^{26}$Al $\rightarrow$ $^{26}$Mg  & 
 $4232.7(2)$  & 46.6 & $493.74(12)_{g_V}(36)_\mu$ & $494.54(12)_{g_V}(2)_\mu$\\
\hline
$^{34}$Cl $\rightarrow$ $^{34}$S  & 
 $5491.66(5)$ & 43.1 & $2051.5(0.5)_{g_V}(1.9)_\mu $ & $2056.0(5)_{g_V}(1)_\mu$ \\
\hline
$^{34}$Ar $\rightarrow$ $^{34}$Cl  & $6061.8(1)$& 42.0 & $3501.3(0.8)_{g_V}(3.5)_\mu $  & $3509.4(8)_{g_V}(2)_\mu$  \\
\hline
$^{38}$K $\rightarrow$ $^{38}$Ar  & 
  $6044.24(5)$ & 41.6 & $3381.1(0.8)_{g_V}(3.6)_\mu$ & $3389.3(8)_{g_V}(3)_\mu$\\
  \hline
$^{38}$Ca $\rightarrow$ $^{38}$K  & 
  $6612.12(7) $ & 41.3  & $5455.6(1.3)_{g_V}(6.0)_\mu$ & $5469.7(1.3)_{g_V}(0.5)_\mu$\\
\hline
$^{42}$Sc $\rightarrow$ $^{42}$Ca & $6426.3(1)$  &  40.3 & $4578.8(1.1)_{g_V}(5.3)_\mu$ & $4591.2(1.1)_{g_V}(0.5)_\mu$\\
\hline
$^{46}$V $\rightarrow$ $^{46}$Ti  & $7052.4(1)$ & 39.2 & $7358.5(1.8)_{g_V}(9.3)_{\mu}$ & $7380.4(1.8)_{g_V}(0.9)_\mu$ 	\\
\hline
$^{50}$Mn $ \rightarrow$ $^{50}$Cr &
$7634.45(7)$ & 38.8 & $10931.6(2.6)_{g_V}(14.9)_{\mu}$ & $10967.2(2.6)_{g_V}(1.5)_\mu$\\
\hline
$^{54}$Co $ \rightarrow$ $^{54}$Fe &
$8244.4(3)$ & 38.3 & $15980.8(3.8)_{g_V} (23.4)_{\mu}$ & $16037.0(3.8)_{g_V} (2.6)_\mu$\\
\hline
           \end{tabular}
    \caption{Combination $\overline{P}$ of the phase space factor, inner and outer radiative corrections for 11 of the superallowed transitions used in the determination of $V_{ud}$~\cite{Hardy:2020qwl}. The second and third columns show the electron-capture $\mathcal Q$ value, $\mathcal Q_{EC}$, and the matching scale $\mu_\pi$. The fourth and fifth column show $\overline{P}$ at $\mathcal O(\alpha)$ and including the $\mathcal O(\alpha^2 Z)$ corrections obtained in this work. The first uncertainty, denoted by the subscript $g_V$, denotes the uncertainty from the single-nucleon vector coupling $g_V$. The second uncertainty, denoted by the subscript $\mu$, is obtained by varying the renormalization scale $\mu_{\text{ext}}$ between $E_0$ and $4 E_0$.     
    }
    \label{tab:exp}
\end{table}

In the formalism of Ref.~\cite{Cirigliano:2024msg}, 
the $\mathcal O(\alpha^2 Z)$ contributions we computed provide a correction to the ``outer corrections''  $\tilde\delta_R' (E_e,\mu)$, which are now given by
\begin{align}\label{eq:drp}
\tilde      \delta_R' (E_e,\bar\mu) = \alpha \, \hat{g}^{(1)}(\beta,\bar E,\bar\mu) + \alpha^2 Z \, \hat{g}^{(2)}(\beta,\bar\mu). 
\end{align}
To give an idea of the size of the $\alpha^2 Z$ corrections, we calculate the combination
\begin{equation}\label{combscale}
\overline{P} =   \left[C_\text{eff}^{(g_V)}\right]^2\bar f (1 + \bar\delta_R^\prime).
\end{equation}
This combination includes the phase space factor, given by 
\begin{equation}\label{eq:PS}
 \bar f(\bar\mu) =  \frac{1}{m_e^5} \int_{m_e}^{E_0} d E_e p_e E_e (E_0 - E_e)^2   \, 
 \bar F (\beta, \bar\mu) \tilde C(E_e),
\end{equation}
the ``inner corrections'', contained in the coefficient $C_{\text{eff}}^{(g_V)}$, and the outer corrections $\bar\delta_R^\prime$. In the phase space factor, we neglect all higher order corrections in $E_e/k_F$, thus retaining only the contributions that arise for point-like nuclei. This is accomplished by setting $\tilde C(E_e) = 1$~\cite{Cirigliano:2024msg}. 
The bar  on $\delta_R^\prime$ in Eq.~\eqref{combscale} denotes the phase-space average, defined as 
\begin{align}
\bar \delta^\prime_R(\bar\mu) = \frac{\int_{m_e}^{E_0} d E_e p_e E_e (E_0 - E_e)^2  
    \, 
 \bar F (\beta, \bar\mu) \,  
    \tilde \delta_R^\prime(E_e,\bar\mu)  }{\int_{m_e}^{E_0} d E_e  p_e E_e (E_0 - E_e)^2\, 
 \bar F (\beta, \bar\mu)  }.
\label{eq:PSaverage}
 \end{align}
The combination in Eq.~\eqref{combscale} is convenient, as it contains minimal nuclear physics input, and thus allow us to assess the importance of perturbative QED corrections.
The half-life is then obtained by combining $\overline{P}$ with the nuclear-structure-dependent corrections $\delta_{\text{NS}}$ and $\delta_C$  as 
\begin{equation}
    \frac{1}{t}  =     \frac{G_F^2 |V_{ud}|^2  m_e^5}{\pi^3 \log 2}\,   \overline{P}
     (1 +  \bar \delta_\text{NS}) \,  ( 1 - \bar  \delta_C).
    \label{eq:master2}
\end{equation}
We stress that the correction to the phase space $\tilde C(E_e)$ is not negligible, and needs to be included in the extraction of $V_{ud}$. However, it does not impact the discussion of the perturbative corrections obtained here.
 
The three objects in Eq.~\eqref{combscale} are scale dependent.
For consistency with the determination of $g_V$ in Ref.~\cite{Cirigliano:2023fnz}, we switch to the $\overline{\text{MS}}_\chi$ scheme by setting
\begin{equation}
    \bar \mu = \bar \mu_\chi e^{-1}.
\end{equation}
We  evaluate $C_{\text{eff}}^{(g_V)}$ at the matching scale $\bar\mu_\chi = \mu_\pi$, given in Eq.~\eqref{eq:matchscale}. We report the value of $\mu_\pi$ for the isotopes of interest in Table~\ref{tab:exp}. 
We then evolve $C_{\text{eff}}^{(g_V)}$ to a low-energy renormalization scale $\bar\mu_\chi = \mu_{\text{ext}}$
using the anomalous dimension in Eq.~\eqref{eq:CeffRG}. 
$C^{(g_V)}_{\text{eff}}$ thus contains fixed order terms of $\mathcal O(\alpha)$, $\mathcal O(\alpha^2 Z^2)$ and $\mathcal O(\alpha^2 Z)$, and the resummation of large logarithms of $\mu_{\text{ext}}/\mu_\pi$ in the series 
\begin{equation}
    \alpha^n L^n,\;  \alpha^n L^{n-1},\; (\alpha^2 Z^2 L)^n,\; (\alpha^2 Z L)^n,\;   \alpha^2 Z^2 \alpha^n L^{n-1},\;    \alpha^2 Z \alpha^n L^{n-1}. 
\end{equation}
We evaluate the Fermi function $\bar{F}(\beta,\bar\mu)$ and the outer corrections at the same low-energy scale. The Fermi function contains terms at all orders in $\alpha^n Z^n$~\cite{Hill:2023acw}. 
We here extend the calculation of $\delta^\prime_R$ to include terms of $\mathcal O(\alpha^2 Z)$.
We choose the central value of the
low-energy renormalization scale to be  
$\mu_{\rm ext} = 2 E_0$. We then vary 
$\mu_{\rm ext}$ between $E_0$ and $4 E_0$ to assess the importance of missing higher order corrections. The values of the endpoint energy for the
11 superallowed transition we consider can be read off Table~\ref{tab:exp}, using 
$E_0 = \mathcal Q_{EC} - m_e$,
where $\mathcal Q_{EC}$ is the electron capture $\mathcal Q$ value, given in Ref.~\cite{Hardy:2020qwl}.
Finally, we notice that $C_{\text{eff}}^{(g_V)}$
depends on the scale $\Lambda$. For this scale we choose the arbitrary value $R_A = 1.2 A^{1/3}$ fm, as done in Ref.~\cite{Cirigliano:2024msg}. In a complete calculation of the decay rate, the $\Lambda$ dependence will be absorbed by nuclear matrix elements of effective operators. 

\begin{figure}
    \centering
    \includegraphics[width=0.8\linewidth]{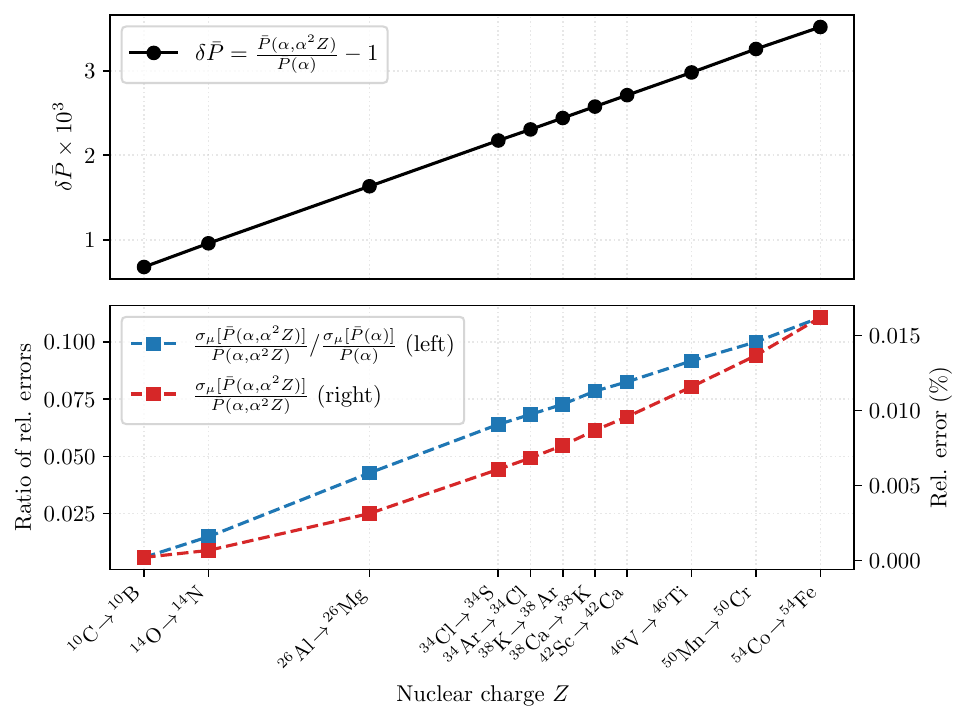}
    \caption{$\delta \overline{P}$ (top panel) and relative uncertainties for different beta decays ($Z$). $\delta \overline{P}$ is expected to show a linear behavior since it encodes the $\mathcal{O}(\alpha^2 Z)$ corrections computed in this work. For the same reason, in the bottom panel, the ratio of relative uncertainties, shown in blue and with scale shown on the left axis,  exhibits a linear behavior. 
    The relative uncertainty on $\overline{P}(\alpha,\alpha^2 Z)$ is shown in red, with scale on the right axis.
    The quadratic dependence of the relative uncertainty of the $\mathcal{O}(\alpha^2Z)$ contribution is due to missing $\mathcal{O}(\alpha^3Z^2)$ terms.}
    \label{fig:plot1}
\end{figure}

The fourth and fifth columns of Table~\ref{tab:exp}
report our evaluation of $\overline{P}$,
including  $\mathcal O(\alpha)$ and $\mathcal O(\alpha^2 Z)$ corrections to $\delta^\prime_R$,
respectively. Both objects contain the $\overline{\text{MS}}$
Fermi function at all orders, which provides the largest electromagnetic correction.
For each entry, we list two sources of uncertainty. The first, indicated by the subscript $g_V$, is the uncertainty induced by the single-nucleon vector coupling $g_V$, given in Eq.~\eqref{eq:gVmu0}. This uncertainty is nucleus-independent and corresponds to a relative uncertainty on $\overline{P}$ of about $2.4 \cdot 10^{-4}$.
The second uncertainty, denoted by the subscript $\mu$, is obtained by varying $\mu_{\text{ext}}$ between $E_0$ and $4E_0$.
As expected, the $\mathcal O(\alpha^2 Z)$ correction grows with $Z$. This is illustrated in Fig.~\ref{fig:plot1}.
Defining 
\begin{equation}
   \delta \overline{P} = \frac{\overline{P}(\alpha, \alpha^2 Z)}{\overline{P}(\alpha)} - 1,
\end{equation}
we see that $\delta \overline{P}$ ranges from
$\delta \overline{P} = 6.9 \cdot 10^{-4}$ for the $^{10}$C $\rightarrow$ $^{10}$B decay,  to $\delta \overline{P} = 3.6 \cdot 10^{-3}$ in the decay of $^{54}$Co. 
We notice that the size of the $\alpha^2 Z$ corrections is underestimated by the scale variation of the $\mathcal O(\alpha)$ result. Indeed, the  difference between 
$\overline{P}(\alpha)$ and $\overline{P}(\alpha, \alpha^2 Z)$
corresponds to about twice the scale variation at $\mathcal O(\alpha)$.  
As illustrated in the bottom panel of Fig.~\ref{fig:plot1},
after inclusion of the $\mathcal O(\alpha^2 Z)$ corrections, the scale variation uncertainty becomes negligible for all nuclei considered here. 
The blue line in Fig.~\ref{fig:plot1}, with scale shown on the left axis, shows the ratio of the scale uncertainty in $\overline{P}(\alpha, \alpha^2 Z)$ and
$\overline{P}(\alpha)$. We see that, even in the most unfavorable case of $^{54}$Co, the scale variation is reduced by a factor of ten.
The residual scale dependence arises from missing  
$\mathcal O(\alpha^2)$ and $\mathcal O(\alpha^3 Z^2)$. 
As a consequence, the uncertainty grows quadratically in $Z$, as shown by the red line in Fig.~\ref{fig:plot1} (scale shown on the right axis). This uncertainty reaches the 10$^{-4}$ level for $A\gtrsim 40$, but it is still subleading with respect to the uncertainty in $g_V$.
Finally,  Ref.~\cite{Plestid:toappear} recently computed $\mathcal O(\alpha^2 Z)$ corrections in the relativistic limit, $\beta \rightarrow 1$. We find that, for the nuclei of interest, the relativistic limit provides a very good approximation of the full result in Eq.~\eqref{eq:dg2}. The relative difference in $\overline{P}$ calculated with Eq.~\eqref{eq:dg2} or with its relativistic limit, Eq.~\eqref{eq:relativistic},  is between $1 \cdot 10^{-4}$ and $2\cdot 10^{-4}$ for all the nuclei in Table~\ref{tab:exp}.

\section{Computational framework}
\label{sec:computational}

\begin{figure}
     \centering
     \begin{subfigure}[b]{0.24\textwidth}
         \centering
         \includegraphics[width=0.6\textwidth]{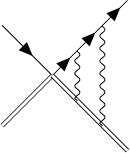}
         \caption{}
         \label{fig:y equals x}
     \end{subfigure}
     \hfill
          \begin{subfigure}[b]{0.24\textwidth}
         \centering
         \includegraphics[width=0.6\textwidth]{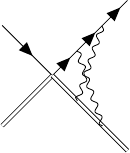}
         \caption{}
         \label{fig:y equals x}
     \end{subfigure}
     \hfill
          \begin{subfigure}[b]{0.24\textwidth}
         \centering
         \includegraphics[width=0.6\textwidth]{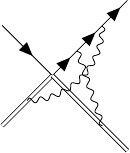}
         \caption{}
         \label{fig:y equals x}
     \end{subfigure}
     \hfill
          \begin{subfigure}[b]{0.24\textwidth}
         \centering
         \includegraphics[width=0.6\textwidth]{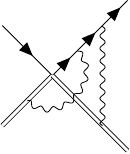}
         \caption{}
         \label{fig:y equals x}
     \end{subfigure}
     \hfill
          \begin{subfigure}[b]{0.24\textwidth}
         \centering
         \includegraphics[width=0.6\textwidth]{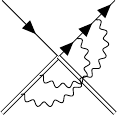}
         \caption{}
         \label{fig:y equals x}
     \end{subfigure}
     \hfill
          \begin{subfigure}[b]{0.24\textwidth}
         \centering
         \includegraphics[width=0.6\textwidth]{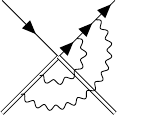}
         \caption{}
         \label{fig:y equals x}
     \end{subfigure}
     \hfill
          \begin{subfigure}[b]{0.24\textwidth}
         \centering
         \includegraphics[width=0.6\textwidth]{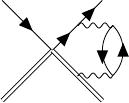}
         \caption{}
         \label{fig:y equals x}
     \end{subfigure}
     \hfill
          \begin{subfigure}[b]{0.24\textwidth}
         \centering
         \includegraphics[width=0.6\textwidth]{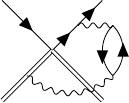}
         \caption{}
         \label{fig:y equals x}
     \end{subfigure}
     \hfill
          \begin{subfigure}[b]{0.24\textwidth}
         \centering
         \includegraphics[width=0.6\textwidth]{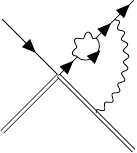}
         \caption{}
         \label{fig:y equals x}
     \end{subfigure}
     \hfill
          \begin{subfigure}[b]{0.24\textwidth}
         \centering
         \includegraphics[width=0.65\textwidth]{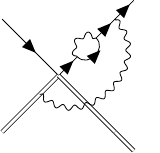}
         \caption{}
         \label{fig:y equals x}
     \end{subfigure}
     \hfill
          \begin{subfigure}[b]{0.24\textwidth}
         \centering
         \includegraphics[width=0.6\textwidth]{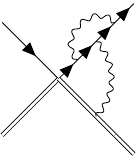}
         \caption{}
         \label{fig:y equals x}
     \end{subfigure}
     \hfill
          \begin{subfigure}[b]{0.24\textwidth}
         \centering
         \includegraphics[width=0.6\textwidth]{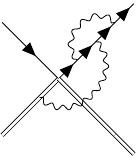}
         \caption{}
         \label{fig:y equals x}
     \end{subfigure}
     \hfill
        \caption{Virtual-virtual diagrams contributing to $\beta$ decay at $\mathcal{O}(\alpha^2Z)$. Double, plain, and wavy lines denote nuclei, leptons and photons, respectively.}
        \label{fig:diagsvirtualvirtual}
\end{figure}

The evaluation of the $\mathcal{O}\left(\alpha^2Z\right)$ corrections involves a relatively large number of multi-loop diagrams with distinct topologies and overlapping infrared and ultraviolet singularities. In this section, we describe the computational strategy and tools used to handle these challenges.
The calculation includes both virtual–virtual and real–virtual diagrams, shown in Figs.~\ref{fig:diagsvirtualvirtual} and~\ref{fig:virtualreal}, which contribute to the ultraviolet and infrared structure of the amplitude. Their interplay ensures the cancellation of divergences and the infrared safety of the total rate.
To compute the diagrams contributing to $\beta$ decay at $\mathcal{O}\left(\alpha^2Z\right)$ we follow standard multi-loop techniques. We generate all relevant Feynman diagrams using \texttt{qgraf}~\cite{Nogueira:1991ex}, perform tensor reduction to express the amplitudes in terms of scalar integrals, and finally use \texttt{Kira}~\cite{Maierh_fer_2018,Klappert:2020nbg,Lange:2025fba,fermat, firefly2, firefly1} to reduce these integrals to a minimal set of master integrals.
Some of the master integrals can be evaluated using the standard procedure: introducing Feynman or $\lambda$ parameters, completing the square, performing the integral(s) over loop momenta and finally a trivial integral over Feynman or $\lambda$ parameters. However, most integrals exhibit a higher degree of complexity. For these, we made use of several techniques~\cite{Smirnov:2012gma} such as the Cheng-Wu theorem~\cite{Cheng:1987ga}, Mellin-Barnes techniques~\cite{Belitsky:2022gba,Czakon:2005rk,Smirnov:2009up,Ochman:2015fho}, sector decomposition~\cite{Binoth:2000ps,Binoth:2003ak} and differential equations~\cite{Remiddi:1997ny,Gehrmann:1999as,Argeri:2007up,Henn:2014qga}. The system of differential equations and the analytic expressions for all master integrals are presented in Appendix~\ref{sec:masterintegrals}.
All integrals are regularized in dimensional regularization, $D=4-2\varepsilon$. Ultraviolet divergences are removed by the counterterms discussed in Appendix~\ref{sec:countertermlagrangian}, while infrared singularities cancel once real and virtual corrections are combined. However, for practical purposes, in this paper we do not distinguish $\varepsilon_{\text{UV}}$ and $\varepsilon_{\text{IR}}$.
\subsection{Virtual-virtual diagrams}

We define the class of two-loop integrals needed for the virtual-virtual calculation as
\begin{align}
I\left(\alpha,\beta,\gamma_1,\gamma_2,\sigma_1,\sigma_2,\sigma_3\right)&=-\left(4\pi\right)^4\mu^{4\varepsilon}\int_{l,k}\frac{1}{\left(v\cdot l\right)^{\alpha}}\frac{1}{\left(v\cdot k\right)^{\beta}}\frac{1}{\left(k^2\right)^{\gamma_1}}\frac{1}{\left(l^2\right)^{\gamma_2}}  \nonumber\\
 &  \times  \frac{1}{\left((k+l)^2+2p_e\cdot (k+l)\right)^{\sigma_1}}\frac{1}{\left(l^2+2p_e\cdot l\right)^{\sigma_2}}\frac{1}{\left(k^2+2p_e\cdot k\right)^{\sigma_3}}\,,
    \label{eq:twoloopint}
\end{align}
where $v$ is the nuclear velocity, $p_e$ denotes the electron four-momentum, $p_e^2 = m_e^2$, and
\begin{equation}
    \int_k \equiv \int \frac{d^d k}{(2\pi)^d}.
\end{equation}
A feature of the integrals in Eq.~\eqref{eq:twoloopint} is that contain linear propagators, as common in heavy particle effective field theories. The reduction techniques used in \texttt{Kira} also apply to linear propagators~\cite{Maierh_fer_2018,Klappert:2020nbg,Lange:2025fba}.

Whenever a certain power has tilde, we substitute the corresponding propagator for a $\delta$-function and add an extra factor of $-(2\pi i)$, e.g.
\begin{align}
I\left(\tilde{1},\beta,\gamma_1,\gamma_2,\sigma_1,\sigma_2,\sigma_3\right)&=\left(4\pi\right)^4\left(2\pi i\right)\mu^{4\varepsilon}\int_{l,k}\delta\left(v\cdot l\right)\frac{1}{\left(v\cdot k\right)^{\beta}}\frac{1}{\left(k^2\right)^{\gamma_1}}\frac{1}{\left(l^2\right)^{\gamma_2}}  \nonumber\\
    &\times \frac{1}{\left((k+l)^2+2p_e\cdot (k+l)\right)^{\sigma_1}}\frac{1}{\left(l^2+2p_e\cdot l\right)^{\sigma_2}}\frac{1}{\left(k^2+2p_e\cdot k\right)^{\sigma_3}}, \nonumber \\
I\left(\tilde{1},\tilde{1},\gamma_1,\gamma_2,\sigma_1,\sigma_2,\sigma_3\right)&=-\left(4\pi\right)^4(2\pi i)^2\mu^{4\varepsilon}\int_{l,k}\delta\left(v\cdot l\right)\delta\left(v\cdot k\right)\frac{1}{\left(k^2\right)^{\gamma_1}}\frac{1}{\left(l^2\right)^{\gamma_2}} \nonumber\\
    &\times \frac{1}{\left((k+l)^2+2p_e\cdot (k+l)\right)^{\sigma_1}}\frac{1}{\left(l^2+2p_e\cdot l\right)^{\sigma_2}}\frac{1}{\left(k^2+2p_e\cdot k\right)^{\sigma_3}}\,.
    \label{eq:twoloopint2}
\end{align}
In intermediate steps of the calculation we have to define extra integral classes such as the one with $k\cdot p_e \to -k\cdot p_e$ or $l \cdot p_e \to - l \cdot p_e$. Nonetheless, at a final stage the different classes combine to form integrals belonging to the class in Eq.~\eqref{eq:twoloopint}, with two, one or no cut heavy particle propagators for contributions at $\mathcal O(\alpha^2 Z^2)$, $\mathcal O(\alpha^2 Z)$ and $\mathcal O(\alpha^2)$, respectively \footnote{The master integral $f_8$ apparently violates this rule. Though it has a heavy particle propagator, $f_8$ is $\beta$-independent, and symmetric under $k \rightarrow -k$.
This can be used to show that $\delta f_8 = 2 f_8$, where  $\delta f_8  \propto I(0,\tilde{1},0,0,1,1,0)$. We can thus use $f_8$ and $\delta f_8$ interchangeably in Eq.~\eqref{eq:alpha2zmasters} }.

The full virtual-virtual calculation at $\mathcal{O}(\alpha^2 Z)$ depends on 12 master integrals 
\begin{align}
    \delta f_1& = \frac{1}{m_e^3} \left(\frac{\bar{\mu}^2}{m_e^2} \right)^{-2\varepsilon} I\left(\tilde{1},0,1,0,1,0,0\right),  \, \,&\delta f_2&= \frac{1}{m_e^4} \left(\frac{\bar\mu^2}{m_e^2} \right)^{-2\varepsilon} I\left(\tilde{1},-1,1,0,1,0,0\right), \nonumber \\
    \delta f_4&= \frac{1}{m_e} \left(\frac{\bar\mu^2}{m_e^2} \right)^{-2\varepsilon} I\left(\tilde{1},0,1,1,1,0,0\right), \, \,&\delta f_5&=\left(\frac{\bar\mu^2}{m_e^2} \right)^{-2\varepsilon} I\left(\tilde{1},1,1,1,1,0,0\right), \nonumber \\
    \delta f_7&= \frac{1}{m_e^3} \left(\frac{\bar\mu^2}{m_e^2} \right)^{-2\varepsilon} I\left(\tilde{1},0,0,0,1,1,0\right),   \, \,& f_8&= \frac{1}{m_e^3} \left(\frac{\bar\mu^2}{m_e^2} \right)^{-2\varepsilon}  I\left(0,1,0,0,1,1,0\right), \nonumber \\
    \delta f_9&= \frac{1}{m_e^2} \left(\frac{\bar\mu^2}{m_e^2} \right)^{-2\varepsilon}  I\left(1,\tilde{1},0,0,1,1,0\right), \,  \,&\tilde{\delta} f_9&= \frac{1}{m_e^2} \left(\frac{\bar\mu^2}{m_e^2}\right)^{-2\varepsilon} I\left(\tilde{1},1,0,0,1,1,0\right), \nonumber \\
    \delta f_{10}&= \frac{1}{m_e} \left(\frac{\bar\mu^2}{m_e^2}\right)^{-2\varepsilon} I\left(0,\tilde{1},0,1,1,1,0\right),  \, \,&\delta f_{11}&= \frac{1}{m_e^2} \left(\frac{\bar\mu^2}{m_e^2}\right)^{-2\varepsilon} I\left(-1,\tilde{1},0,1,1,1,0\right), \nonumber \\
    \delta f_{14}&= \frac{1}{m_e} \left(\frac{\bar\mu^2}{m_e^2}\right)^{-2\varepsilon} I\left(0,\tilde{1},0,0,1,1,1\right),   \, \,&\delta f_{15}&= \frac{1}{m_e^3} \left(\frac{\bar\mu^2}{m_e^2}\right)^{-2\varepsilon} I\left(0,\tilde{1},-1,0,1,1,1\right)\,,
    \label{eq:alpha2zmasters}
\end{align}
where $\bar\mu$ is the $\overline{\text{MS}}$ renormalization scale
\begin{equation}
     \mu = \bar{\mu} \left(\frac{e^{\gamma_E}}{4\pi}\right)^\frac{1}{2}.
\end{equation}
The dimensionless functions in Eq.~\eqref{eq:alpha2zmasters} are functions of $\beta$ alone. 
The IBP reduction leads to factors of $1/\varepsilon$ multiplying the integrals, meaning that we have to compute the master integrals to higher orders in $\varepsilon$. The master integrals that we need to $\mathcal{O}(\varepsilon^2)$ are $\delta f_1\left(\beta\right)$, $\delta f_2\left(\beta\right)$, $\delta f_7\left(\beta\right)$, $\delta f_9\left(\beta\right)$, $\tilde{\delta} f_9\left(\beta\right)$, $\delta f_{10}$ and $\delta f_{11}\left(\beta\right)$. The ones needed to $\mathcal{O}(\varepsilon)$ are $\delta f_4\left(\beta\right)$, $\delta f_5\left(\beta\right)$,  $f_8\left(\beta\right)$, $\delta f_{14}\left(\beta\right)$ and $\delta f_{15}\left(\beta\right)$.
 For completeness, we also report the master integrals needed for the $\mathcal{O}(\alpha^2 Z^2)$ calculation: 
 \begin{align}
    \delta\delta  f_5\left(\beta\right)&=  \left(\frac{\bar\mu^2}{m_e^2}\right)^{-2\varepsilon}I\left(\tilde{1},\tilde{1},1,1,1,0,0\right), \,\quad &\delta\delta  f_9\left(\beta\right)&= \frac{1}{m_e^2}\left(\frac{\bar\mu^2}{m_e^2}\right)^{-2\varepsilon} I\left(\tilde{1},\tilde{1},0,1,1,0,0\right) \,.
    \label{eq:z2masters}
\end{align}

We can trivially observe the increase in complexity when going from $\mathcal{O}(\alpha^2 Z^2)$ to $\mathcal{O}(\alpha^2 Z)$; we find more masters, and more complicated masters. This becomes even more accentuated if we look at the master integrals relevant for the $\mathcal{O}(\alpha^2)$ virtual-virtual calculation. 
In this case, we would need to calculate the functions $f_{1,2}(\beta)$,
$f_{4,5}(\beta)$, $f_{7}(\beta)$, $f_9(\beta)$, $f_{10,11}(\beta)$ and $f_{14,15}(\beta)$ that are obtained from the corresponding $\delta f$ in Eq.~\eqref{eq:alpha2zmasters} without cutting a heavy particle propagator, and are needed at the same order in $\varepsilon$ as the cut integrals. Moreover, we would have to consider three additional master integrals
\begin{align}
    f_3(\beta) &= \frac{1}{m_e^2} \left(\frac{\bar\mu^2}{m_e^2}\right)^{-2\varepsilon} I\left(0,0,1,1,1,0,0\right), \nonumber \\
    f_6(\beta) &= \frac{1}{m_e^4} \left(\frac{\bar\mu^2}{m_e^2}\right)^{-2\varepsilon} I\left(0,0,0,0,1,1,0\right), \nonumber\\
    f_{13}(\beta) &= \frac{1}{m_e^2} \left(\frac{\bar\mu^2}{m_e^2}\right)^{-2\varepsilon} I\left(0,0,0,0,1,1,1\right), \label{eq:alpha2masters}
\end{align}
which are needed at $\mathcal{O}(\varepsilon^2)$. The functions in Eq.~\eqref{eq:alpha2masters} are standard relativistic integrals. They do not appear in the $\mathcal O(\alpha^2 Z)$ calculation, since they do not contain any heavy particle propagator that can be cut.

\subsection{Real-virtual diagrams}\label{sub:real}
\begin{figure}
     \centering
     \begin{subfigure}[b]{0.19\textwidth}
         \centering
         \includegraphics[width=0.5\textwidth]{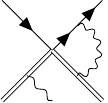}
         \label{fig:five over x}
     \end{subfigure}
          \hfill
     \begin{subfigure}[b]{0.19\textwidth}
         \centering
         \includegraphics[width=0.5\textwidth]{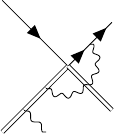}
         \label{fig:five over x}
     \end{subfigure}
          \hfill
     \begin{subfigure}[b]{0.19\textwidth}
         \centering
         \includegraphics[width=0.5\textwidth]{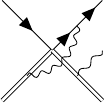}
         \label{fig:five over x}
     \end{subfigure}
          \hfill
     \begin{subfigure}[b]{0.19\textwidth}
         \centering
         \includegraphics[width=0.5\textwidth]{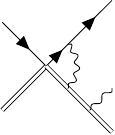}
         \label{fig:five over x}
     \end{subfigure}
          \hfill
     \begin{subfigure}[b]{0.19\textwidth}
         \centering
         \includegraphics[width=0.5\textwidth]{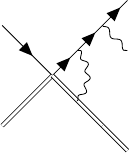}
         \label{fig:five over x}
     \end{subfigure}
          \hfill
     \begin{subfigure}[b]{0.19\textwidth}
         \centering
         \includegraphics[width=0.5\textwidth]{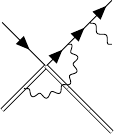}
         \label{fig:five over x}
     \end{subfigure}
          \hfill
     \begin{subfigure}[b]{0.19\textwidth}
         \centering
         \includegraphics[width=0.5\textwidth]{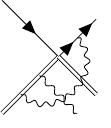}
         \label{fig:five over x}
     \end{subfigure}
          \hfill
     \begin{subfigure}[b]{0.19\textwidth}
         \centering
         \includegraphics[width=0.5\textwidth]{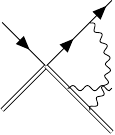}
         \label{fig:five over x}
     \end{subfigure}
          \hfill
     \begin{subfigure}[b]{0.19\textwidth}
         \centering
         \includegraphics[width=0.5\textwidth]{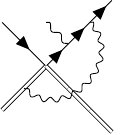}
         \label{fig:five over x}
     \end{subfigure}
          \hfill
     \begin{subfigure}[b]{0.19\textwidth}
         \centering
         \includegraphics[width=0.5\textwidth]{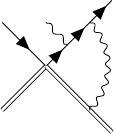}
         \label{fig:five over x}
     \end{subfigure}
          \hfill
        \caption{Virtual-real diagrams. The notation is as in Fig.~\ref{fig:diagsvirtualvirtual}.}
        \label{fig:virtualreal}
\end{figure}

At $\mathcal{O}(\alpha^2 Z^2)$ and $\mathcal{O}(\alpha^2 Z)$ we do not find any contribution coming from real-real emissions, as sketched in:
\begin{align}
    \raisebox{-0.4\height}{\includegraphics[width=0.12\columnwidth]{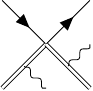}}+\raisebox{-0.4\height}{\includegraphics[width=0.12\columnwidth]{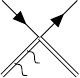}}+\raisebox{-0.4\height}{\includegraphics[width=0.12\columnwidth]{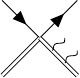}}\sim Z^2+(Z-1)^2-2Z(Z-1)=1\,,
    \label{eq:realreal}
\end{align}
and an analogous argument holds for photons emitted from the electron line.
Therefore, for the goal of this paper we can directly move to real-virtual emissions, for which we show the Feynman diagrams in Fig.~\ref{fig:virtualreal}. If we denote the photon momenta by $\ell$ and the loop momenta by $k$, we define the integral class by 
\begin{align}
I^{\text{RV}}\left(\tilde{1},\beta,\gamma,\delta\right)=-2\pi (4\pi)^2\mu^{2\varepsilon}\int_k\delta\left(v\cdot k\right)\frac{1}{\left(k^2\right)^\beta} \frac{1}{\left((k+p_e)^2-m_e^2\right)^\gamma} \frac{1}{\left((k+p_e+\ell)^2-m_e^2\right)^\delta} \,.
\end{align}
After performing tensor and IBP reduction, we obtain 5 master integrals

\begin{align}
    \delta g_2& = \frac{1}{m_e} \left(\frac{\bar{\mu}^2}{m_e^2} \right)^{-\varepsilon} I^{RV}\left(\tilde{1},0,1,0\right),  \, \,&\delta g_3&= \frac{1}{m_e} \left(\frac{\bar\mu^2}{m_e^2} \right)^{-\varepsilon} I^{RV}\left(\tilde{1},0,0,1\right), \nonumber \\
    \delta g_5&= m_e \left(\frac{\bar\mu^2}{m_e^2} \right)^{-\varepsilon} I^{RV}\left(\tilde{1},1,0,1\right), \, \,&\delta g_6&=m_e\left(\frac{\bar\mu^2}{m_e^2} \right)^{-\varepsilon} I^{RV}\left(\tilde{1},0,1,1\right), \nonumber \\
    \delta g_8&= m_e^3\left(\frac{\bar\mu^2}{m_e^2} \right)^{-\varepsilon} I^{RV}\left(\tilde{1},1,1,1\right)\nonumber\,. \\
    \label{eq:realvirtualmasters}
\end{align}
The only master that we need at $\mathcal{O}(\varepsilon)$ is $\delta g_2$, all the others are only needed at $\mathcal{O}(\varepsilon^0)$.
We calculate these master integrals in the soft limit.
We find that both $\delta g_5$ 
and $\delta g_8$ generate terms in the real matrix element squared that go as $(\ell \cdot p_e)^{-2 -2 \varepsilon}$. These however cancel when combined, so that the 
real matrix element scales like $|\mathcal R|^2 \rightarrow \ell^{-2}$ in the soft limit. 
We can thus use  standard one-loop techniques to treat  the integration over the photon phase space~\cite{Frixione:1995ms}.
Namely, we write the photon phase space  as~\cite{Alioli:2010xd}
\begin{align}
  \mu^{2\varepsilon}\int  \frac{d^{d-1} \ell}{(2\pi)^{d-1} 2 \ell} |\mathcal R(\ell) |^2 &= \frac{2^{2-2\varepsilon}}{(4\pi)^{2}} \frac{\Gamma(1-\varepsilon)}{\Gamma(1-2\varepsilon)} \left(\frac{\bar{\mu}^2 e^{\gamma_E} } {\bar E^2}\right)^{\varepsilon}  \nonumber \\ & \times  \int_0^1 d \xi  \frac{\theta(\xi)}{\xi^{1+2 \varepsilon}}  \frac{1}{2}\int_{-1}^1  d y (1-y^2)^{-\varepsilon}  \frac{1}{\pi} \, \int_0^\pi d \phi  (\sin \phi)^{-2 \varepsilon} ( \xi^2 |\mathcal R(\xi)|^2),
\end{align}
where we rescaled the photon momentum $|\vec{\ell}|=\bar E \, \xi$, with $\xi \in (0,1)$.
As $\xi^2 |\mathcal R(\xi)|^2$ tends to a constant, we can expand $\xi^{-1-2\varepsilon}$ in $\varepsilon$ as~\cite{Frixione:1995ms}
\begin{align}
    \frac{1}{\xi^{1+2\varepsilon}} = - \frac{1}{2\varepsilon} \delta(\xi)  + \left[\frac{1}{\xi}\right]_{+} + \mathcal O(\varepsilon),
\end{align}
with the plus distribution defined as 
\begin{equation}
  \int d\xi  \left[\frac{1}{\xi}\right]_{+}  f(\xi) \equiv \int_0^1 d\xi \frac{1}{\xi}  \left[f(\xi) -f(0) \right].
\end{equation}
To evaluate the pole part, we only need the soft limit $|\vec{\ell}|\ll m_e$ of the master integrals, which we give in Eq.~\eqref{eq:mastersrealvirtualsolution}.
To evaluate the contribution from the plus distribution part, we realize that the coefficients of the master integrals $\delta g_{3,5,6}$ are real and of $\mathcal{O}(\varepsilon^0)$. However, the integrals $\delta g_{3,5,6}$ are imaginary at $\mathcal{O}(\varepsilon^0)$, implying that they do not contribute once we take the real part of the result. The coefficient of $\delta g_8$ is also real and of $\mathcal{O}(\varepsilon^0)$, but we can split $\delta g_8=\delta h_8+\delta g_8|_{\text{soft}}$, where we find $\delta h_8$ to be finite and imaginary, but $\delta g_8|_{\text{soft}}$ contains a real finite part. Therefore, we only need the results for $\delta g_2=\delta g_2|_{\text{soft}}$ and $\delta g_8|_{\text{soft}}$, which can be found in Eq.~\eqref{eq:mastersrealvirtualsolution}.

\section{Conclusion}
\label{sec:conclusion}

We have performed the first complete calculation of the $\mathcal{O}(\alpha^2 Z)$ radiative corrections to superallowed $0^+\to 0^+$ Fermi $\beta$ decays in a low-energy heavy-particle effective field theory that describes the interactions of ultrasoft photons with nuclei. This theory is the last level in a tower of effective field theories  that rigorously separates the contributions of hard, soft, 
potential and ultrasoft photon modes~\cite{Cirigliano:2024msg,Cirigliano:2024rfk}, allowing for the systematic improvement of the calculation of the various objects that enter  the factorized decay rate,  and for the resummation of large logarithms of ratios of the energy scales affecting these process. 
All virtual-virtual and real-virtual diagrams have been computed analytically after reducing the integrals to a set of master integrals, which we evaluated 
using various methods, including Mellin-Barnes techniques, sector decomposition and differential equations.
Our main results are encoded in the functions $\hat{g}^{(2)}\left(\beta,\bar E,\bar \mu\right)$ and $\hat{a}^{(2)}\left(\beta,\bar E,\bar \mu\right)$, 
given in Eqs.~\eqref{eq:dg2}
and \eqref{eq:da2}, which determine the full $\mathcal{O}(\alpha^2 Z)$ contribution to the outer radiative correction $\tilde{\delta}^\prime_R\left(E_e,\bar \mu\right)$.

When combined with the known $\mathcal{O}(\alpha)$ and $\mathcal{O}(\alpha^2 Z^2)$ terms, and with nuclear-structure-independent terms in the matching coefficient $C_{\text{eff}}^{(g_V)}$, the new corrections modify the phase space factor $\overline{P}$ by up to $4\times 10^{-3}$ for the heaviest superallowed transition we considered, $^{54}$Co $\rightarrow ^{54}$Fe. The inclusion of these corrections significantly reduces the residual renormalization scale dependence of the decay rate, stabilizing the perturbative expansion and confirming the consistency of the EFT framework. 
The residual scale variation indicates that missing $\mathcal O(\alpha^2)$
and $\mathcal O(\alpha^3 Z^2)$ will impact the decay rate at the $10^{-4}$ level.

For an extraction of $V_{ud}$, the ultrasoft corrections computed here need to be combined with the single-nucleon vector coupling, $g_V$, which contains nonperturbative information on the $\gamma W$ box, and with nuclear-structure-dependent corrections.
The latter include $\mathcal O(\alpha^2)$ terms in the nuclear matrix elements of the operators $\mathcal V^{}_+$ and $\mathcal V^{\text{3b}}_+$, which cancel the dependence on the scale $\Lambda$ implicit in $\overline{P}$, and provide an \textit{ab initio} definition of the  
nuclear radius $R$ that has been traditionally included in the Fermi function and in the outer corrections. 
The combination of the ultrasoft loops we computed, of $C^{(g_V)}_{\text{eff}}$, given in Ref.~\cite{Cirigliano:2024msg}, 
and of the nuclear matrix elements of $\mathcal V^{}_+$ and $\mathcal V^{\text{3b}}_+$ will provide the complete $\mathcal O(\alpha^2 Z)$ corrections to the decay rate, and will be important for a robust extraction of $V_{ud}$.

Extensions of this work include the calculation of the $\mathcal{O}(\alpha^2)$ corrections without $Z$-enhancement, or the $Z^2$-enhanced $\mathcal{O}(\alpha^3Z^2)$ corrections. As we mentioned, the residual scale variations seem to indicate that their effects is below the current level of precision. However, the former will also impact neutron decay~\cite{VanderGriend:2025mdc,Cao:2025lrw}, for which 
new precise measurements of the lifetime and decay correlations are planned~\cite{Alarcon:2023gfu}. The latter grow with $Z$, and it is important to check that their impact is not underestimated by the scale variations.
Finally, we can extend the heavy particle Lagrangian to higher orders in $E_e R$, to include, for example, the effects of nuclear radii. The techniques developed in this work can be straightforwardly used to calculate QED corrections at higher orders in $E_e R$, and to account for effects that are now included phenomenologically in the correction to the phase space $\tilde C_e$.

\paragraph{Note added.} Near the completion of this work, the authors of Ref.~\cite{Plestid:toappear} informed us of their own independent calculation of $\mathcal O(\alpha^2 Z)$ correction to superallowed $\beta$ decays in the relativistic limit, $\beta \rightarrow 1$. The relativistic limit of our results fully agrees with Ref. \cite{Plestid:toappear}. As the calculations were completely independent and carried out with different techniques, this agreement
provides an important check of both results.

\acknowledgments
We thank Fabian Lange for his help regarding the use of \texttt{Kira}.
We thank J. Kumar for discussions and comments, and for his involvement in the initial stages of the project.
We acknowledge A.~Vicini for several insightful discussion on multi-loop methods. 
We thank R. Plestid, Z.~Cao, R.~Hill and P.~Vander Griend  for sharing their $\mathcal O(\alpha^2 Z)$ results and for discussions. We are grateful to
V.~Cirigliano, W.~Dekens, J.~de~Vries, M.~Hoferichter and P.~Stoffer for their encouragement, mentorship and for comments on the manuscript. We thank C.~Y.~Seng for comments on the manuscript. 
\`OLC gratefully acknowledges financial support by the Swiss National Science Foundation (Project No. PCEFP2\_194272 and mobility grant PCEFP2\_194272/3) and a UZH Candoc Grant (Grant No. [FK-25-094]).
Financial support by Los Alamos
National Laboratory's Laboratory Directed Research and Development program under projects 20250164ER  and 
20260246ER
is gratefully acknowledged.  Los Alamos National Laboratory is operated by Triad National Security, LLC,
for the National Nuclear Security Administration of U.S.\ Department of Energy (Contract No.\
89233218CNA000001). 
We acknowledge support from the DOE Topical Collaboration ``Nuclear Theory for New Physics,'' award No.\ DE-SC0023663. 

\appendix

\section{Renormalization}
\label{sec:countertermlagrangian}

To renormalize the theory, we split the Lagrangian into interaction and counterterm pieces \cite{Collins:1984xc}. The interaction Lagrangian is given by
\begin{align}\label{eq:appLag1}
    \mathcal L &=  
    - \frac{1}{4}   F^{\mu \nu} F_{\mu \nu} + \bar \nu i \slashed{\partial} \nu +
    \bar e  \left( i \slashed{\partial} - m_e\right)  e    +  {\mathcal A}^\dagger_f\, \left( i v\cdot \partial + \Delta \right) {\mathcal A}_f +  {\mathcal A}^\dagger_i\, i v\cdot \partial {\mathcal A}_i- e Q_e \mu^{\varepsilon} \bar e \slashed{A} e
\nonumber \\ &  
  - e \mu^\varepsilon v \cdot A \left(  Z {\mathcal A}^\dagger_f \mathcal A_f + (Z-1) {\mathcal A}^\dagger_i \mathcal A_i \right)  
- \frac{2 G_F}{\sqrt{2}} V_{ud} \,  \mu^{2\varepsilon} \left(  C_V {\mathcal A}^\dagger_f v_\mu {\mathcal A}_i \, \bar e \gamma^\mu P_L \nu + {\rm h.c.}\right), 
\end{align}
where we focus here on electron emission, the positron case being analogous. 
The counterterm Lagrangian is 
\begin{align}\label{eq:ct}
    \mathcal L_{ct} &=   {\mathcal A}^\dagger_f \delta Z_{A_f}  \left[
    i v \cdot D + \Delta
    \right] \mathcal A_f
    + {\mathcal A}^\dagger_i \delta Z_{A_i}      i v \cdot D  \mathcal A_i
    - \frac{2 G_F}{\sqrt{2}} V_{ud} \delta C_V 
    \mathcal A^\dagger_f v_\mu \mathcal A_i
    \bar e \gamma^\mu P_L\nu \nonumber \\
   & + \delta Z_2 \bar e i \slashed{D}  e -  \delta Z_m m_e \bar e e - \frac{1}{4}\delta Z_3 F^{\mu \nu} F_{\mu \nu}, 
\end{align}
where we used the QED relation between bare and renormalized charges  and
\begin{equation}
    \delta C_V =  C_{V0} \left(Z_{A_i} Z_{A_f} Z_e \right)^{\frac{1}{2}} - C_V \mu^{2\varepsilon}, \qquad \delta Z_m =  Z_2 Z_m - 1.
\end{equation}
We then define the $\overline{\text{MS}}$ coupling $C_V$ by subtracting $1/\varepsilon$ poles in the two-point functions and   $\mu^{2\varepsilon} \times [1/\varepsilon]$ poles in the calculation of the four-point function $\mathcal A_i \rightarrow \mathcal A_f e \nu$ \cite{Collins:1984xc}.    
Here we define the couplings in $\overline{\text{MS}}$, while we use the pole mass definition of the electron mass.
The one-loop counterterms are then given by \cite{Manohar:2000dt}
\begin{align}
    \delta Z_{A_f} &= \frac{\alpha}{4\pi} (Z)^2 \frac{2}{\varepsilon_{\rm UV}}, \\
        \delta Z_{A_i} &= \frac{\alpha}{4\pi} (Z - 1)^2 \frac{2}{\varepsilon_{\rm UV}}, \\
    \delta C_{V} &= \mu^{2\varepsilon} C_V \left(  -\frac{\alpha}{4\pi}  \frac{1}{\varepsilon_{\rm UV}} + \frac{\alpha}{4\pi} (Z - 1) Z \frac{2}{\varepsilon_{\rm UV}}  \right),  \label{eq:a7}\\
    \delta Z_2 &= - \frac{\alpha}{4\pi}   \frac{1}{\varepsilon_{\rm UV} }, \\
    \delta Z_3 &= - \frac{4}{3} n_\ell \frac{\alpha}{4\pi}   \frac{1}{\varepsilon_{\rm UV} },
\\
  \delta Z_m &= -
 \frac{\alpha}{4\pi}  
\left(  \frac{3}{\varepsilon_{\rm UV} } + 3 L_\mu + 4 \right). \label{eq:ZM}
\end{align}
$n_\ell$ is the number of charged leptons.
As scales $\mu_{\text{ext}} \ll k_F$, only the electron is active, and we can take $n_\ell = 1$.
With this definition of $\delta Z_m$,  $m_e$ denotes the position of the pole of the electron propagator, up to $\mathcal O(\alpha^2)$ corrections \cite{Tarrach:1980up}. 
The residue at the poles of the electron and nuclear propagators at $\mathcal O(\alpha)$ is given by
\begin{align}
  R_e &= 1- \frac{\alpha}{4\pi}  \left(    \frac{2}{\varepsilon_{\rm IR}} +4 + 3 \log \frac{\mu^2}{m_e^2}\right), \\
  R_{A_i} &= 1 + \frac{\alpha}{4\pi} (Z - 1)^2\left(  - \frac{2}{\varepsilon_{\rm IR}}\right),\\
R_{A_f} &= 1 + \frac{\alpha}{4\pi} (Z)^2\left(  - \frac{2}{\varepsilon_{\rm IR}}\right).
\end{align}
Notice that 
that for the heavy baryon field 
\begin{equation}
    R_{A_i} = Z^{-1}_{A_i}, \qquad R_{A_f} = Z^{-1}_{A_f}.
\end{equation}
This is true at all orders in $\alpha$ when using $\overline{\text{MS}}$ to regulate both infrared and ultraviolet divergences, as all the loops vanish, and the renormalized propagator is equal to the counterterm.
Using the one-loop determination of $\delta C_V$ to solve for the bare coupling, we find
\begin{equation}
    C_{V 0} =  C_V \mu^{2\varepsilon}  \left[1 + \frac{\alpha}{4\pi\varepsilon} \left( - 1 + 2 Z (Z-1)\right)  \right](Z_{A_i} Z_{A_f} Z_e)^{-\frac{1}{2}},
\end{equation}
and, exploiting the fact that $C_{V0}$ is $\mu$-independent
and $d \alpha/d\log \mu = - 2 \varepsilon \alpha + \ldots$, we obtain the lowest order anomalous dimension as
\begin{equation}
    \frac{d}{d\log \mu} C_V = -\frac{\alpha}{\pi} \frac{3}{4} C_V,
\end{equation}
in agreement with Eq.~\eqref{eq:anomalous}.

At $\mathcal O(\alpha^2 Z)$, we can avoid the calculation of the heavy field wavefunction renormalization. 
After adding real and virtual diagrams, but before including the two-loop wavefunction renormalization, we find that  the squared amplitude has a local  divergence, of the form
\begin{equation}
\frac{\mathcal A^{(2)}  \mathcal A^{(0) *} +\mathcal A^{(0)}  \mathcal A^{(2) *}  }{2 E_e E_\nu} =  \alpha^2 C_V^2 \mu^{4\varepsilon} \left[ \frac{Z(Z - 1)}{4\varepsilon} \right]. 
\end{equation}
This divergence is absorbed by the counterterm diagram
\begin{equation}
  \left. \left( C_V \mu^{2\varepsilon} + \delta C_V\right)^2 R_{A_i} R_{A_f}
     - C_V^2 \mu^{4\varepsilon} \right|_{\alpha^2} = - \alpha^2 C_V^2 \mu^{4\varepsilon} \frac{Z (Z- 1)}{4\varepsilon},
\end{equation}
or
\begin{equation}\label{eq:cv2}
    C_V \mu^{2\varepsilon} + \delta C_V = C_V \mu^{2\varepsilon} \left[1 - \frac{\alpha^2 Z (Z - 1)}{8\varepsilon} \right] \left[R_{A_i} R_{A_f}\right]^{-\frac{1}{2}},
\end{equation}
where we neglected the $\mathcal O(\alpha)$ term.
To get the running of $C_V$,
we can plug Eq.~\eqref{eq:cv2}  into the expression for the bare coupling.
At $\mathcal O(\alpha^2 Z)$, we can neglect the electron wavefunction renormalization,
and, using $R_{A_{i,f}} = Z^{-1}_{A_{i,f}}$, we see that the factors of the residues drop out.
\begin{equation}
    C_{V0} = C_V \mu^{2\varepsilon} \left[1 - \frac{\alpha^2 Z (Z-1)}{8\varepsilon} \right] + \mathcal O(\alpha^2).
\end{equation}
Taking the derivative with respect to $\mu$ we then get
\begin{equation}
    \frac{d}{d\log \mu} C_V = - \frac{\alpha^2}{2}  Z(Z-1),
\end{equation}
which reproduces the $\mathcal O(\alpha^2 Z)$ expansion of the anomalous dimension given in Eq.~\eqref{eq:anomalous}
and derived in Refs. \cite{Hill:2023acw,Borah:2024ghn,Cirigliano:2024msg}

\section{Master Integrals}
\label{sec:masterintegrals}

\subsection{Differential Equations}
The derivatives $\partial_\beta$ and $\partial_m$ in  terms of the derivative $\partial/\partial p_e^\mu$ read
\begin{align}
    \frac{\partial}{\partial \beta}&=\frac{1}{\beta(1-\beta^2)}p_e^\mu\frac{\partial}{\partial p_e^\mu}-\frac{m_e}{\beta \sqrt{1-\beta^2}}v^\mu \frac{\partial}{\partial p_e^\mu} \nonumber \\
    \frac{\partial}{\partial m_e}&=\frac{1}{m_e}p_e^\mu \frac{\partial}{\partial p_e^\mu},
\end{align}
where we used the constraint $p_e^2 = m_e^2$.
The differential equations for the master integrals contributing to the virtual-virtual calculation at $\mathcal{O}(\alpha^2Z)$, as defined in Eq.~\eqref{eq:alpha2zmasters}, are
\begingroup
\allowdisplaybreaks[4]
\begin{align}
    \partial_\beta \delta f_1 &=\frac{5-6\varepsilon}{\beta(1-\beta^2)}\delta f_1 +4\frac{1-\varepsilon}{\beta \sqrt{1-\beta^2}}\delta f_2 \nonumber \\
    \partial_\beta \delta f_2 &=\frac{2\beta(-1+\varepsilon)}{(1-\beta^2)^{3/2}}\delta f_1 \nonumber \\
    \partial_\beta \delta f_4 &=\frac{2\varepsilon-1}{\beta(1-\beta^2)}\delta f_4+\frac{4\sqrt{1-\beta^2}(-1+\varepsilon)}{\beta^3}\delta f_2+\frac{-5+6\varepsilon}{\beta^3}\delta f_1 \nonumber \\
    \partial_\beta \delta f_5 &=\frac{-4\varepsilon}{\beta(1-\beta^2)}\delta f_5-2\frac{1-5\varepsilon+6\varepsilon^2}{\beta\sqrt{1-\beta^2}(-1+4\varepsilon)}\delta f_4 -6\frac{(1-\beta^2)(1-3\varepsilon+2\varepsilon^2)}{\beta^5(-1+4\varepsilon)}\delta f_2 +\nonumber \\&+\frac{\sqrt{1-\beta^2}(-1+2\varepsilon)(15-18\varepsilon+\beta^2(-3+4\varepsilon))}{2\beta^5(-1+4\varepsilon)}\delta f_1 \nonumber \\
    \partial_\beta \delta f_7 &=\frac{1-2\varepsilon}{\beta(1-\beta^2)}\delta f_7 \nonumber \\
    \partial_\beta  f_8 &=0 \nonumber \\
    \partial_\beta \delta f_9 &=\frac{2-4\varepsilon}{\beta(1-\beta^2)}\delta f_9+\frac{6-8\varepsilon}{\beta\sqrt{1-\beta^2}}f_8 \nonumber \\
    \partial_\beta \tilde \delta f_9 &=\frac{2-4\varepsilon}{\beta(1-\beta^2)}\tilde \delta f_9 +\frac{2-2\varepsilon}{\beta\sqrt{1-\beta^2}}\delta f_7 \nonumber \\
    \partial_\beta \delta f_{10} &=\frac{2-4\varepsilon}{\beta(1-\beta^2)}\delta f_{10}+\frac{3-4\varepsilon}{\beta\sqrt{1-\beta^2}}\delta f_{11}+2\frac{\sqrt{1-\beta^2}(-1+\varepsilon)}{\beta^3}\delta f_2 +\frac{-5+6\varepsilon}{2\beta^3}\delta f_1 \nonumber \\
    \partial_\beta \delta f_{11} &=\frac{-1+2\varepsilon}{\beta(1-\beta^2)}\delta f_{11}+\frac{\beta(-1+2\varepsilon)}{(1-\beta^2)^{3/2}}\delta f_{10}+\frac{1-2\varepsilon}{2\beta\sqrt{1-\beta^2}}\delta f_1 +\frac{-1+2\varepsilon}{\beta\sqrt{1-\beta^2}}f_8 \nonumber \\
    \partial_\beta \delta f_{14} &=\frac{-\beta^2-2\varepsilon+2\beta^2\varepsilon}{\beta(1-\beta^2)}\delta f_{14}+\frac{-2+3\varepsilon}{2\beta}\delta f_{15}+\frac{1-\varepsilon}{\beta}\delta f_7  + \frac{1-2\varepsilon}{2\beta}f_8 \nonumber \\
    \partial_\beta \delta f_{15} &=\frac{-1+4\beta^2+2\varepsilon-6\beta^2\varepsilon}{\beta(1-\beta^2)}\delta f_{15}+\frac{4\beta(-1+2\varepsilon)}{1-\beta^2}\delta f_{14}+\frac{4\beta(-1+\varepsilon)}{1-\beta^2}\delta f_7 +\frac{2-4\varepsilon}{\beta}f_8 \nonumber \\ \label{eq:differential}
\end{align}
\endgroup
while the ones for the master integrals that contribute to the virtual-virtual calculation at $\mathcal{O}(\alpha^2 Z^2)$ (defined in Eq.~\eqref{eq:z2masters}) read
\begin{align}
    \partial_{\beta} \delta\delta f_5&=-\frac{4\varepsilon}{\beta(1-\beta^2)}\delta \delta f_5 \nonumber \\
    \partial_\beta \delta\delta f_9 &=\frac{2-4\varepsilon}{\beta(1-\beta^2)}\delta\delta f_9.
\end{align}
Most of the  differential equations in Eq.~\eqref{eq:differential} are decoupled.
To simplify the differential equations for $\delta f_{10}$ and $\delta f_{11}$, we introduce 
\begin{equation}
    \delta h_{11}(\beta) =  \delta f_{11}(\beta) + \frac{1}{\sqrt{1-\beta^2}} \delta f_{10}(\beta).
\end{equation}
In terms of $\delta f_{10}$ and $\delta h_{11}$, the system becomes 
\begin{align}
\partial_\beta \delta f_{10}(\beta) &= - \frac{1}{\beta (1-\beta^2)} \delta f_{10}(\beta)   + \frac{3-4\varepsilon}{\beta \sqrt{1-\beta^2}} \delta h_{11}(\beta)  - \frac{5-6\varepsilon}{2\beta^3}\delta f_1(\beta) \nonumber \\ &  - (1-\varepsilon)\frac{2\sqrt{1-\beta^2}}{\beta^3} \delta f_2(\beta) 
\\
    \partial_\beta \delta h_{11}(\beta) & = + \frac{2(1-\varepsilon)}{\beta(1-\beta^2)} \delta h_{11}(\beta)  -  \frac{2 \varepsilon}{\beta\sqrt{1-\beta^2}} \delta f_{10}(\beta) - \frac{1-2\varepsilon}{\beta\sqrt{1-\beta^2}} f_8  \nonumber \\
    & + \frac{- 5 + 6 \varepsilon + \beta^2 (1-2\varepsilon)}{2\beta^3 \sqrt{1-\beta^2}} \delta f_1(\beta) - \frac{2(1-\varepsilon)}{\beta^3} \delta f_2(\beta),
\end{align}
where, in the second equation, the coupling between $\delta h_{11}$ and $\delta f_{10}$ now starts at $\mathcal O(\varepsilon)$.
Similarly, a partial decoupling between $\delta f_{14}$
and $\delta f_{15}$ is obtained by using $    f_{\pm} =\delta f_{14} \pm \delta f_{15}$.
The presence of $\sqrt{1-\beta^2}$ in the differential equations \eqref{eq:differential}
is partially an artifact of our choice of defining dimensionless functions by factoring out powers of the electron mass in Eq.~\eqref{eq:alpha2zmasters}.
We could have equally well have defined dimensionless functions by pulling out powers of the electron energy. Defining $\delta\varphi_i$ as in Eq.~\eqref{eq:alpha2zmasters}, but with $m_e$ replaced by $E_e$, we find the following differential equations 
\begingroup
\allowdisplaybreaks[1]
\begin{align}
    \partial_\beta \delta \varphi_1 &=\frac{5-6\varepsilon + \beta^2 (-3+4\varepsilon) }{\beta( 1-\beta^2)}\, \delta \varphi_1 +\frac{4(1-\varepsilon)}{\beta(1-\beta^2)}\, \delta \varphi_2, \nonumber \\
    \partial_\beta \delta \varphi_2 &= 
    - \frac{2\beta}{1-\beta^2}  (1-\varepsilon) \left(2  \delta \varphi_2 + \delta \varphi_1 \right)
    , \nonumber \\
        \partial_\beta \delta \varphi_4 &=\ 
        \frac{1}{\beta(1-\beta^2)} \left[ \left(-1+ 2\varepsilon + \beta^2 (-1 + 4 \varepsilon)\right) \, \delta \varphi_4  + \frac{-5+6\varepsilon}{\beta^2} \, \delta \varphi_1 - \frac{4 (1-\varepsilon)}{\beta^2} \, \delta \varphi_2\right]
        \nonumber \\
    \partial_\beta \delta \varphi_5 &=
    \frac{1}{\beta(1-\beta^2)} \left[-4\varepsilon (1-\beta^2) \, \delta \varphi_5
    + \frac{1-2\varepsilon}{1-4\varepsilon} \left( \frac{15 -18 \varepsilon + \beta^2 (-3+4 \varepsilon)}{2\beta^4} \, \delta \varphi_1 \right. \right. \nonumber \\ & \left. \left.  + \frac{6 (1-\varepsilon)}{\beta^4} \, \delta \varphi_2  + 2 (1-3\varepsilon) \, \delta \varphi_4\right)
     \right]
    \nonumber \\
    \partial_\beta \delta \varphi_7 &= \frac{1 - 2 \varepsilon + \beta^2 (-3+4\varepsilon)}{\beta(1-\beta^2)} \,\delta \varphi_7 \nonumber \\
    \partial_\beta  \varphi_8 &= - \frac{\beta (3-4\varepsilon)}{1-\beta^2}\,  \varphi_8 \nonumber \\
    \partial_\beta \delta \varphi_9 &= \frac{2}{\beta(1-\beta^2)} \left[ (1-\beta^2)(1-2\varepsilon) \, \delta \varphi_9  + (3-4\varepsilon) \,  \varphi_8 \right] \nonumber \\
    \partial_\beta \tilde \delta \varphi_9 &= \frac{2}{\beta( 1-\beta^2)} \left[ (1-\beta^2)(1-2\varepsilon) \, \tilde\delta \varphi_9 + (1-\varepsilon)\delta \varphi_7 \right] \nonumber \\
    \partial_\beta \delta \varphi_{10} &= \frac{2}{\beta(1-\beta^2)} \left[ \left( (1-2\varepsilon) + \frac{\beta^2}{2}( -1+4\varepsilon) \right) \delta \varphi_{10} + \left(\frac{3}{2} - 2 \varepsilon\right) \delta \varphi_{11} - \frac{5-6\varepsilon}{4\beta^2} \delta\varphi_1 \right.  \nonumber \\ & \left. - \frac{1-\varepsilon}{\beta^2} \delta\varphi_2\right] \nonumber \\
    \partial_\beta \delta \varphi_{11}  &= \frac{2}{\beta(1-\beta^2)} \left[\frac{1}{2} (1+2\beta^2) (-1+2\varepsilon) \, \delta \varphi_{11} + \frac{\beta^2}{2} (-1+2\varepsilon) \, \delta \varphi_{10}
    + \frac{1}{4} (1-2\varepsilon) \, \delta \varphi_1 \right. \nonumber \\   &\left. + \left(-\frac{1}{2} + \varepsilon\right)\,  \varphi_8
    \right] \nonumber \\
    \partial_\beta \delta \varphi_{14} &=
    \frac{2}{\beta(1-\beta^2)} \left[ \left(-\varepsilon + \beta^2 (-1+3\varepsilon)
    \right) \delta\varphi_{14} 
     - \frac{1}{4} (2-3\varepsilon) \delta \varphi_{15} + \frac{1-\varepsilon}{2} \delta\varphi_7    + \frac{1}{4} (1-2\varepsilon) \varphi_8
     \right]\nonumber \\
    \partial_\beta \delta \varphi_{15}&=
    \frac{2}{\beta(1-\beta^2)}  \left[
    -\frac{1}{2}\left(1-\beta^2\right)(1-2\varepsilon)
    \delta\varphi_{15} - 2 \beta^2 (1-\beta^2) (1-2\varepsilon)\delta\varphi_{14} \nonumber \right. \\ & \left. - 2\beta^2 (1-\varepsilon)  \delta\varphi_7 + (1-\beta^2)(1-2\varepsilon)\, \varphi_8
    \right].  \label{eq:differentialEe}
\end{align}

\subsection{Analytic expressions of the master integrals}
In this section we will report the analytic results for the real part of the master integrals that contributed to our calculation. We find that the expressions have a slightly more compact form expressed in terms of the variable $y=(1-\beta)/(1+\beta)$. 
\begingroup
\allowdisplaybreaks[4]
\begin{align}
    \delta f_1(\beta)&=\frac{2 \pi^2 (1+y)}{\sqrt{y}}
    \Bigg\{ 1
        +\varepsilon  \left[3 \log y + \frac{1+18y-9y^2 -2 y^3}{3 y (1+y)} \right]  +\varepsilon^2  \left[  18 \PolyLog(y) + \frac{7}{2}\log^2 y +  \frac{\pi^2}{6} \nonumber \right. \\ &\left. + 3 \frac{-1-9y+9y^2 + y^3}{y (1+y)}  \log(1-y) + \frac{2+45 y - 18 y^2 - 5 y^3}{3 y(1+y)} \log y  \nonumber \right. \\ & \left. -  \frac{2(-11-126 y+ 171 y^2+ 22 y^3)}{9y (1+y)}\right] \Bigg\},\nonumber \\
    \delta f_2(\beta)&= - \frac{\pi^2}{y} \Bigg\{ 
  1 + 3 y + y^2 + \varepsilon \left[3 (1 + 3 y + y^2) \log y  + \frac{1+46 y+ 27 y^2-38 y^3 - 2 y^4}{6 y} \right]  \nonumber \\
  & + \varepsilon^2 \left[ (1+3y + y^2) \left(18 \PolyLog(y) + \frac{7}{2}\log^2 y +  \frac{\pi^2}{6} \right) - \frac{3 (1-y^2) (1+28 y+y^2)}{2y} \log(1-y) \nonumber \right. \\ & \left. + \frac{2 + 110 y+ 81 y^2- 86 y^3 - 5 y^4}{6y} \log y + \frac{22+ 715 y -216 y^2 -1079 y^3 - 44 y^4}{18 y}  \right]
    \Bigg\}, \nonumber \\
\delta f_4(\beta) &=   \frac{4\pi^2 \sqrt{y}}{1-y} \Bigg\{ \log y + \varepsilon \left[ 2 \PolyLog(y) - \frac{\pi^2}{3} + \log^2 y  + 2 \log(1-y)\log y  + \log y \nonumber \right. \\ & \left.  + \frac{-1+3y-2y^2}{y}\right] \Bigg\}, \nonumber \\
\delta f_5(\beta) &=  -2 \pi^2 \Bigg\{ \frac{1}{\varepsilon} +  6 + 2 \frac{1+y}{1-y}\log y 
+ \varepsilon \left[ 8 \frac{2-y}{1-y} \left(\PolyLog(y) + \log y \log(1-y)\right) + \frac{3-11y}{1-y} \frac{\pi^2}{6} \right. \nonumber \\ & \left. + 2 \frac{1+y}{1-y} \log^2 y  + 12 \frac{1+y}{1-y} \log y  + 36 \right]
\Bigg\}, \nonumber
\\
\delta f_7(\beta) &= - \frac{2\pi^2 (1-y)}{\sqrt{y}} \Bigg\{ 1 + \varepsilon \left[3 + \log y - 2 \log (1-y)\right]
+ \varepsilon^2 \left[\frac{1}{2} (\log y -2 \log(1-y))^2 \right. \nonumber \\ & \left. 
+ 3 \log y -6\log (1-y) + \frac{\pi^2}{6} + 7 
\right]
\Bigg\}, \nonumber \\
f_8 &= \frac{32}{3} \pi^2 \varepsilon + \frac{64}{9} \pi^2  \varepsilon^2 (11 -12 \log(2)), \nonumber \\
 \delta f_9(\beta)&= - 2 \pi^2 \frac{(1-y)^2}{y}-\varepsilon \frac{4 \pi^2}{y} \left[2 (1-y)^2 + 4 \sqrt{y} (1+y) + (1-y)^2 \left( \log y - 4 \log (1+\sqrt{y}) \right)\right] \nonumber \\
 &+\varepsilon^2 \frac{4\pi^2}{y} \left\{ 4(1-y)^2 \left[4 \textrm{Li}_2(\sqrt{y})
- 4 \textrm{Li}_2\left(\frac{1+\sqrt{y}}{2}\right) - {\rm Li}_2(y) - 2 \log^2 (1+\sqrt{y})
\nonumber \right. \right. \\ & \left. \left.- 4 \log (1+\sqrt{y}) \log (1-\sqrt{y}) + \log(1-y)\log y - \frac{1}{4}\log^2(y) - 2\log^2(2) 
\nonumber \right. \right. \\
& \left. \left. 
+ 4 (1-\log(2)) \log(1+\sqrt{y})
-  \log(y) + 4 \log(1-y)\log(2) + \frac{\pi^2}{48}
\right] \nonumber \right. \\ & \left. 
+ 32 \sqrt{y} (1+y) \log(2) 
+ 2 (-3 + 6 y - 3 y^2 - 16 \sqrt{y} (1+y))\right\} \nonumber \\
     \tilde \delta f_9(\beta)&=2\pi^2 \frac{1-y^2}{y}+\varepsilon \frac{4\pi^2 (1-y)}{y} \left( \log(y) - (1+y)\log(1-y) + 2(1+y)\right)\nonumber \\
 &-\varepsilon^2 \frac{4\pi^2}{y} (1-y) \left\{ 
 (1-y) \left[ {\rm Li}_2(-y) +
{\rm Li}_2\left(-\frac{1-y}{1+y}\right) - {\rm Li}_2\left(\frac{1-y}{1+y}\right) + \log y \log(1+y) \right] \nonumber \right. \\ & \left. + \frac{1}{2} (-2+y) \log^2(y) -  (1+y) \log^2(1-y) + 2 \log y \log(1-y) - \frac{\pi^2}{6} y \right.  \nonumber \\ & \left.
+ 4 (1+y) \log(1-y) -4 \log y -6 (1+y)
\right\} \nonumber \\
 \delta f_{10}(\beta)&= \frac{\pi^2}{\sqrt{y} (1-y)} \left(1 - y^2 + 4 y \log y \right) +\varepsilon \frac{4\pi^2 \sqrt{y}}{1-y} \Bigg\{
    6 {\rm Li}_2(y) - 8 {\rm Li}_2(\sqrt{y}) \nonumber  \\ &  + \frac{3}{2} \log^2 y + 2 \log y \log(1-y)   - \frac{2(1-y^2)}{y} \log (1+\sqrt{y})  + \frac{3 + 4 y - 7 y^2}{4y} \log y  \nonumber  \\ &  + \frac{\pi^2}{3} +\frac{3}{2y} + \frac{4 - 4 y - 3 y \sqrt{y}}{2 \sqrt{y}}
\Bigg\}
\nonumber \\
 \delta h_{11}(\beta)&\coloneqq \delta f_{11}(\beta)+\frac{1}{\sqrt{1-\beta^2}}\delta f_{10}(\beta)= 
 \frac{\pi^2}{3y} (1-y)^2
 \nonumber \\
 &+\varepsilon\frac{4\pi^2}{y}
\left[ - \frac{2}{3} (1-y)^2 \log (1+ \sqrt{y})  + \frac{1}{12 (1-y)} (3- 9 y +21 y^2 - 7 y^3) \log y \nonumber \right. \\ & \left.  + \frac{19}{36} (1-y)^2 + \frac{1}{3} \sqrt{y} (2 + \sqrt{y} + 2y)
\right]\nonumber \\ & +\varepsilon^2\frac{(1-y)^2}{y}\ 4\pi^2 \frac{782 + 69 \pi^2 + 252 \log(2) + 36 \log^2(2)}{216}  +
\varepsilon^2\frac{4\pi^2}{y} \times \nonumber \\
&  \left\{
(1-y)^2 \Bigg[2 {\rm Li}_2\left(\frac{1+\sqrt{y}}{2}\right)
- \frac{2}{3}
{\rm Li}_2\left(\frac{1-\sqrt{y}}{2}
\right) + \frac{1}{2} \log^2{2} + \frac{2}{3} \log^2 (1+\sqrt{y}) \nonumber \right.  \\ &  \left.
- \frac{2}{3} \log^2 (1-\sqrt{y})  + \frac{2}{3} \log (1-\sqrt{y}) \log (1+\sqrt{y}) 
+ \frac{2}{3} \log(2) \log (1 + \sqrt{y}) \nonumber \right. \\ & \left. -2 \log(2) \log(1-\sqrt{y}) 
\Bigg]     
+ \frac{4}{3(1-y)} (-1 + 3 y - 9 y^2 + 3 y^3) 
{\rm Li}_2 (\sqrt{y}) \nonumber \right. \\ & \left.  +\frac{1-3y + 39 y^2 - 13 y^3 }{6(1-y)} {\rm Li}_2 (y) 
+ \frac{7-21 y+ 57 y^2 - 19 y^3}{24 (1-y)}\log^2 y 
\right. \nonumber \\ & \left.
+ \frac{2 (-3 + 9 y - 6 y^2+ 2y^3)}{3(1-y)} \log y \log(1-y)
 + \frac{2 (1-3 y + 3 y^2-y^3)}{3(1-y)}
\log^2 (1-y) \nonumber \right. \\ & \left. + \frac{-38+90 y-90y^2+38 y^3}{9 (1-y) } \log (1+ \sqrt{y})   +\frac{57-171 y+ 279 y^2-133 y^3}{36 (1-y)} \log y \right. \nonumber \\ & \left. + \frac{-7+21 y-21 y^2+ 7 y^3 -32 \sqrt{y} (1-y^2)}{6(1-y)}\log(2)  - \frac{\pi^2}{36 (1-y)} (13-39y + 27 y^2-9 y^3) \right. \nonumber \\
& \left. - \frac{1}{36(1-y)} \left(39 -205 y  + 205 y^2 -39 y^3 - 200 \sqrt{y} (1-y^2)  \right)
 \right\} \nonumber \\
  \delta f_{+}(\beta)&\coloneqq \delta f_{14}(\beta)+\delta f_{15}(\beta)=\frac{2\pi^2}{(-1+y)y^{3/2}}\left((1-y)^2(-1+2y)-4y^2\log(y)\right) \nonumber \\
 &+\varepsilon\frac{4\pi^2}{y^{3/2}} \left\{
    \frac{4 y^2}{1-y} \left(- 4 {\rm Li}_2(\sqrt{y}) + 3 {\rm Li}_2(y) + 2{\rm Li}_2(-y) + \frac{3}{4}\log^2 y + \log y \log(1-y)\right)  \right. \nonumber \\ 
    & \left.
- (1+y)(1-y)^2 \log(1+y)
 + (-1 + y (3-2y)) \log(1-y)  \nonumber \right. \\ & \left.
 + \frac{1}{2(1-y)}\left(
 2 - 6 y + 11 y^2 + 2 y^3 - 2 y^4\right)\log y
 - (1+y)(1-y)^2 \frac{1}{2} \log \frac{1+\sqrt{y}}{1-\sqrt{y}}\right. \nonumber \\ & \left.
 + \frac{4}{3} \frac{y^2}{1-y} \pi^2 
 + \frac{1}{12} \left(27 - 81 y + 48 y^2 + \sqrt{y} (12 - 8 y + 12 y^2)\right)
 \right\}\nonumber \\
  \delta f_{-}(\beta)&\coloneqq \delta f_{14}(\beta)-\delta f_{15}(\beta)=-\frac{2\pi^2}{y^{3/2}}\left(1+y+2y^2\right) +\varepsilon \frac{4\pi^2}{y^{3/2}} \times \nonumber \\
  & \left\{ (1+y)^3 \log(1+y) + (1+y + 2 y^2) \log(1-y) + \frac{-2 -2y+ y^2+ 6 y^3 + 2y^4}{2 (1-y)} \log y \nonumber \right. \\ &
    \left.  + \frac{(1+y)^3}{2} \log \frac{1+\sqrt{y}}{1-\sqrt{y}}
   + \frac{1}{12} \left(-27 - 39 y - 48 y^2 + \sqrt{y} (-12 -40 y - 12 y^2)\right)
    \right\}.
\end{align}

The results for the master integrals contributing to the virtual-virtual calculation at $\mathcal{O}(\alpha^2 Z^2)$ are
\begin{align}
    \delta\delta f_5(\beta)&=-\frac{4\pi^2}{\varepsilon}+8\pi^2\left(-3+\log\left(\frac{4\beta^2}{1-\beta^2}\right)\right)\nonumber \\ &+\varepsilon\frac{2\pi^2}{3}\left(-216+13\pi^2+72\log\left(\frac{4\beta^2}{1-\beta^2}\right)-12\log^2\left(\frac{4\beta^2}{1-\beta^2}\right)\right) \nonumber \\
    \delta\delta f_9(\beta)&=-\frac{16\pi^2\beta^2}{1-\beta^2}+\varepsilon\frac{32\pi^2\beta^2(-2+\log\left(\frac{4\beta^2}{1-\beta^2}\right))}{1-\beta^2}\nonumber \\
    &+\varepsilon^2\frac{8\pi^2\beta^2}{1-\beta^2}\left(3(-8+\pi^2)+16\log\left(\frac{4\beta^2}{1-\beta^2}\right)-4\log^2\left(\frac{4\beta^2}{1-\beta^2}\right)\right).
\end{align}

In the soft limit, the master integrals needed for the real-virtual calculation are given by
\begin{align}
    \left. \delta g_2 \right|_{\text{soft}}  &=     \left. \delta g_3 \right|_{\text{soft}}   = 2\sqrt{\pi} e^{\varepsilon \gamma_E} \,\Gamma\left(-\frac{1}{2}+\varepsilon\right) \left(\frac{-\beta^2 - i \eta}{1-\beta^2}\right)^{\frac{1}{2} -\varepsilon}, \nonumber \\
        \left. \delta g_5 \right|_{\text{soft}} &= 2 \sqrt{\pi} e^{\varepsilon \gamma_E} \left(\frac{-\beta^2-i \eta}{1-\beta^2}\right)^{-\frac{1}{2} -\varepsilon} \left( -\left(\frac{E_e \beta}{2 p_e \cdot \ell }\right)^{2\varepsilon} \Gamma(2\varepsilon) \Gamma\left(\frac{1}{2}- \varepsilon\right) + \frac{1}{2\varepsilon} \Gamma\left(\frac{1}{2} + \varepsilon\right) \right), \nonumber \\
\left. \delta g_6 \right| _{\text{soft}} &=
   2\sqrt{\pi} e^{\varepsilon \gamma_E} \Gamma\left(\frac{1}{2} + \varepsilon\right) \left(\frac{-\beta^2-i \eta}{1-\beta^2}\right)^{-\frac{1}{2} -\varepsilon}, \nonumber \\
\left. \delta g_8 \right|_{\text{soft}} &=   2 \sqrt{\pi} e^{\varepsilon \gamma_E} \left(\frac{-\beta^2-i \eta}{1-\beta^2}\right)^{-\frac{1}{2} -\varepsilon} \frac{E_e^2 (1-\beta^2)}{2 p_e \cdot \ell} \left(\frac{E_e \beta}{2 p_e \cdot \ell}\right)^{2\varepsilon} \Gamma(2\varepsilon) \Gamma\left(\frac{1}{2} - \varepsilon\right),
\end{align}
where $\eta >0$ indicates the causal prescription arising from the propagators.
As we remarked in Section~\ref{sub:real}, the coefficients of the master integrals $\delta g_5$ and $\delta g_8$ are such that the terms going as $(p_e \cdot \ell)^{-2\varepsilon}$ cancel out.  
The decay rate then only depends on the real part of the master integrals, that are given by
\begin{align}
    \mathfrak{Re}\left( \delta g_2|_{\text{soft}}\right)&=-\varepsilon\frac{4\pi^2\beta}{\sqrt{1-\beta^2}} \nonumber \\
    \mathfrak{Re}\left(  \delta g_3|_{\text{soft}}\right)&= 0+\mathcal{O}\left(\varepsilon\right)\nonumber \\
  \mathfrak{Re}\left(   \delta g_5|_{\text{soft}}\right)&=0+\mathcal{O}\left(\varepsilon\right) \nonumber \\
  \mathfrak{Re}\left(   \delta g_6|_{\text{soft}}\right)&=0+\mathcal{O}\left(\varepsilon\right)\nonumber \\
   \mathfrak{Re}\left(  \delta g_8|_{\text{soft}}\right)&=-\frac{m_e^2\pi^2\sqrt{1-\beta^2}}{2\beta \,\ell\cdot p_e}. \nonumber \\
    \label{eq:mastersrealvirtualsolution}
\end{align}

\section{Differential equations for the $\mathcal O(\alpha^2)$ master integrals }
\label{app:alpha2}
We provide here the differential equations for the master integrals that contribute at $\mathcal O(\alpha^2)$, without $Z$ enhancement. We find

\begingroup
\allowdisplaybreaks[1]
\begin{align}
    \partial_\beta  \varphi_1 &=\frac{5-6\varepsilon + \beta^2 (-3+4\varepsilon) }{\beta( 1-\beta^2)}\,  \varphi_1 +\frac{4(1-\varepsilon)}{\beta(1-\beta^2)}\,  \varphi_2, \nonumber \\
    \partial_\beta  \varphi_2 &= 
    - \frac{2\beta}{1-\beta^2}  (1-\varepsilon) \left(2   \varphi_2 +  \varphi_1 \right)
    , \nonumber \\
\partial_\beta \varphi_3 &= - \frac{2 \beta}{1-\beta^2} (1-2\varepsilon)\varphi_3 \nonumber \\
        \partial_\beta  \varphi_4 &=\ 
        \frac{1}{\beta(1-\beta^2)} \Bigg[ \left(-1+ 2\varepsilon + \beta^2 (-1 + 4 \varepsilon)\right) \,  \varphi_4  + \frac{-5+6\varepsilon}{\beta^2} \,  \varphi_1 - \frac{4 (1-\varepsilon)}{\beta^2} \,  \varphi_2  \nonumber \\ &
        - (2-3\varepsilon) \varphi_3
        \Bigg]
        \nonumber \\
    \partial_\beta  \varphi_5 &=
    \frac{1}{\beta(1-\beta^2)} \left[-4\varepsilon (1-\beta^2) \,  \varphi_5
    +  \frac{1-2\varepsilon}{1-4\varepsilon} \left( \frac{15 -18 \varepsilon + \beta^2 (-3+4 \varepsilon)}{\beta^4} \,  \varphi_1 \right. \right. \nonumber \\ & \left. \left.  + \frac{12 (1-\varepsilon)}{\beta^4} \,  \varphi_2  + 4 (1-3\varepsilon) \,  \varphi_4\right)
     \right]
    \nonumber \\
\partial_\beta \varphi_6 &= - \frac{4 \beta (1-\varepsilon)}{1-\beta^2} \varphi_6  \nonumber\\
    \partial_\beta  \varphi_7 &= \frac{1 - 2 \varepsilon + \beta^2 (-3+4\varepsilon)}{\beta(1-\beta^2)} \,\varphi_7 +  \frac{2(1-\varepsilon)}{\beta (1-\beta^2)} \varphi_6 \nonumber \\
    \partial_\beta  \varphi_8 &= - \frac{\beta (3-4\varepsilon)}{1-\beta^2}\,  \varphi_8 \nonumber \\
    \partial_\beta  \varphi_9 &= \frac{2}{\beta(1-\beta^2)} \left[ (1-\beta^2)(1-2\varepsilon) \, \delta \varphi_9 +(1-\varepsilon)\varphi_7
    + \frac{3-4\varepsilon}{2} \,  \varphi_8 \right] \nonumber \\
    \partial_\beta  \varphi_{10} &= \frac{2}{\beta(1-\beta^2)} \left[ \left( (1-2\varepsilon) + \frac{\beta^2}{2}( -1+4\varepsilon) \right)  \varphi_{10} + \left(\frac{3}{2} - 2 \varepsilon\right)  \varphi_{11} - \frac{5-6\varepsilon}{4\beta^2} \varphi_1 \right.  \nonumber \\ & \left. - \frac{1-\varepsilon}{\beta^2} \varphi_2 + \frac{(1-\varepsilon)^2}{1-2\varepsilon} \frac{1}{1-\beta^2} \varphi_6 \right] \nonumber \\
    \partial_\beta  \varphi_{11}  &= \frac{2}{\beta(1-\beta^2)} \left[\frac{1}{2} (1+2\beta^2) (-1+2\varepsilon) \,  \varphi_{11} + \frac{\beta^2}{2} (-1+2\varepsilon) \,  \varphi_{10}
    + \frac{1}{4} (1-2\varepsilon) \, \varphi_1 \right. \nonumber \\   &\left. + \left(-\frac{1}{4} +\frac{ \varepsilon}{2}\right)\,  \varphi_8
    \right] \nonumber \\
    \partial_\beta \varphi_{13} &=- \frac{2\beta}{1-\beta^2} (1-2\varepsilon) \varphi_{13}
\nonumber \\
    \partial_\beta  \varphi_{14} &=
    \frac{2}{\beta(1-\beta^2)} \Bigg[ \left(-\varepsilon + \beta^2 (-1+3\varepsilon)
    \right) \varphi_{14} 
     - \frac{1}{4} (2-3\varepsilon)  \varphi_{15} + \frac{1-\varepsilon}{2} \varphi_7 \nonumber \\ &    + \frac{1}{8} (1-2\varepsilon) \varphi_8 - \frac{(1-\varepsilon)^2}{2(1-2\varepsilon)} \frac{1}{1-\beta^2} \varphi_6
     \Bigg]\nonumber \\
    \partial_\beta  \varphi_{15}&=
    \frac{2}{\beta(1-\beta^2)}  \left[
    -\frac{1}{2}\left(1-\beta^2\right)(1-2\varepsilon)
    \varphi_{15} - 2 \beta^2 (1-\beta^2) (1-2\varepsilon)\varphi_{14} \nonumber \right. \\ & \left. - 2\beta^2 (1-\varepsilon)  \varphi_7 + (1-\beta^2)(1-2\varepsilon)\, \left(\frac{1}{2} \varphi_8 - \varphi_{13}\right)
    \right].  \label{eq:difffull}
\end{align}
The differential equations in Eq.~\eqref{eq:difffull} are very similar to the corresponding equations for the functions with one cut heavy particle propagator in Eq.~\eqref{eq:differentialEe}, with the main difference that now the master integrals $\varphi_3$, $\varphi_6$ and $\varphi_{13}$ also contribute. The functions in Eq.~\eqref{eq:difffull} are typically more divergent than those in Eq.~\eqref{eq:differential}Ee.
 For example, $\varphi_1(\beta)$ and $\varphi_2(\beta)$ start at $\mathcal O(\varepsilon^{-2})$,
and appear in the diagrams in Fig.~\ref{fig:diagsvirtualvirtual} with  coefficients of $\mathcal O(\varepsilon^{-2})$.
The $\mathcal O(\varepsilon)$ and $\mathcal O(\varepsilon^2)$ terms in $\varphi_1$ and $\varphi_2$ lead to the appearance of harmonic polylogarithms of weight 3 and 4 in the $\mathcal O(\alpha^2)$ result.

\bibliographystyle{JHEP} 
\bibliography{biblio}

@article{Cirigliano:2024msg,
    author = "Cirigliano, Vincenzo and Dekens, Wouter and de Vries, Jordy and Gandolfi, Stefano and Hoferichter, Martin and Mereghetti, Emanuele",
    title = "{Ab initio electroweak corrections to superallowed {\ensuremath{\beta}} decays and their impact on Vud}",
    eprint = "2405.18464",
    archivePrefix = "arXiv",
    primaryClass = "nucl-th",
    reportNumber = "INT-PUB-24-021, LA-UR-24-25160",
    doi = "10.1103/PhysRevC.110.055502",
    journal = "Phys. Rev. C",
    volume = "110",
    number = "5",
    pages = "055502",
    year = "2024"
}

@article{Maierh_fer_2018,
   title={Kira—A Feynman integral reduction program},
   volume={230},
   ISSN={0010-4655},
   url={http://dx.doi.org/10.1016/j.cpc.2018.04.012},
   DOI={10.1016/j.cpc.2018.04.012},
   journal={Computer Physics Communications},
   publisher={Elsevier BV},
   author={Maierhöfer, P. and Usovitsch, J. and Uwer, P.},
   year={2018},
   month=sep, pages={99–112} }

@article{Klappert:2020nbg,
    author = {Klappert, Jonas and Lange, Fabian and Maierh{\"o}fer, Philipp and Usovitsch, Johann},
    title = "{Integral reduction with Kira 2.0 and finite field methods}",
    eprint = "2008.06494",
    archivePrefix = "arXiv",
    primaryClass = "hep-ph",
    reportNumber = "TTK-20-24, P3H-20-041, FR-PHENO-2020-11, MITP/20-044",
    doi = "10.1016/j.cpc.2021.108024",
    journal = "Comput. Phys. Commun.",
    volume = "266",
    pages = "108024",
    year = "2021"
}

@article{Nogueira:1991ex,
    author = "Nogueira, Paulo",
    title = "{Automatic Feynman Graph Generation}",
    reportNumber = "IFM-7-91",
    doi = "10.1006/jcph.1993.1074",
    journal = "J. Comput. Phys.",
    volume = "105",
    pages = "279--289",
    year = "1993"
}

@book{Cheng:1987ga,
    author = "Cheng, Hung and Wu, T. T.",
    title = "{Expanding protons: scattering at high-energies}",
    year = "1987"
}

@article{Smirnov:2009up,
    author = "Smirnov, A. V. and Smirnov, V. A.",
    title = "{On the Resolution of Singularities of Multiple Mellin-Barnes Integrals}",
    eprint = "0901.0386",
    archivePrefix = "arXiv",
    primaryClass = "hep-ph",
    doi = "10.1140/epjc/s10052-009-1039-6",
    journal = "Eur. Phys. J. C",
    volume = "62",
    pages = "445--449",
    year = "2009"
}

@article{Binoth:2000ps,
    author = "Binoth, T. and Heinrich, G.",
    title = "{An automatized algorithm to compute infrared divergent multi-loop integrals}",
    eprint = "hep-ph/0004013",
    archivePrefix = "arXiv",
    reportNumber = "LAPTH-789-00, LPT-ORSAY-00-37",
    doi = "10.1016/S0550-3213(00)00429-6",
    journal = "Nucl. Phys. B",
    volume = "585",
    pages = "741--759",
    year = "2000"
}

@article{Binoth:2003ak,
    author = "Binoth, T. and Heinrich, G.",
    title = "{Numerical evaluation of multi-loop integrals by sector decomposition}",
    eprint = "hep-ph/0305234",
    archivePrefix = "arXiv",
    reportNumber = "EDINBURGH-2003-06, IPPP-03-28, DCPT-03-56",
    doi = "10.1016/j.nuclphysb.2003.12.023",
    journal = "Nucl. Phys. B",
    volume = "680",
    pages = "375--388",
    year = "2004"
}

@article{Ochman:2015fho,
    author = "Ochman, Michal and Riemann, Tord",
    title = "{MBsums - a Mathematica package for the representation of Mellin-Barnes integrals by multiple sums}",
    eprint = "1511.01323",
    archivePrefix = "arXiv",
    primaryClass = "hep-ph",
    reportNumber = "DESY-15-209",
    doi = "10.5506/APhysPolB.46.2117",
    journal = "Acta Phys. Polon. B",
    volume = "46",
    number = "11",
    pages = "2117",
    year = "2015"
}

@article{Czakon:2005rk,
    author = "Czakon, M.",
    title = "{Automatized analytic continuation of Mellin-Barnes integrals}",
    eprint = "hep-ph/0511200",
    archivePrefix = "arXiv",
    reportNumber = "WUE-ITP-2005-014",
    doi = "10.1016/j.cpc.2006.07.002",
    journal = "Comput. Phys. Commun.",
    volume = "175",
    pages = "559--571",
    year = "2006"
}

@article{Belitsky:2022gba,
    author = "Belitsky, A. V. and Smirnov, A. V. and Smirnov, V. A.",
    title = "{MB tools reloaded}",
    eprint = "2211.00009",
    archivePrefix = "arXiv",
    primaryClass = "hep-ph",
    doi = "10.1016/j.nuclphysb.2022.116067",
    journal = "Nucl. Phys. B",
    volume = "986",
    pages = "116067",
    year = "2023"
}

@article{Lange:2025fba,
    author = "Lange, Fabian and Usovitsch, Johann and Wu, Zihao",
    title = "{Kira 3: integral reduction with efficient seeding and optimized equation selection}",
    eprint = "2505.20197",
    archivePrefix = "arXiv",
    primaryClass = "hep-ph",
    reportNumber = "ZU-TH 39/25, HU-EP-25/17-RTG",
    month = "5",
    year = "2025"
}

@article{Hill:2023acw,
    author = "Hill, Richard J. and Plestid, Ryan",
    title = "{Field Theory of the Fermi Function}",
    eprint = "2309.07343",
    archivePrefix = "arXiv",
    primaryClass = "hep-ph",
    reportNumber = "CALT-TH/2023-029, FERMILAB-PUB-23-453-T",
    doi = "10.1103/PhysRevLett.133.021803",
    journal = "Phys. Rev. Lett.",
    volume = "133",
    number = "2",
    pages = "021803",
    year = "2024"
}

@article{Hill:2023bfh,
    author = "Hill, Richard J. and Plestid, Ryan",
    title = "{All orders factorization and the Coulomb problem}",
    eprint = "2309.15929",
    archivePrefix = "arXiv",
    primaryClass = "hep-ph",
    reportNumber = "CALT-TH-2023-034, FERMILAB-PUB-23-454-T",
    doi = "10.1103/PhysRevD.109.056006",
    journal = "Phys. Rev. D",
    volume = "109",
    number = "5",
    pages = "056006",
    year = "2024"
}

@article{Wilkinson:1982hu,
    author = "Wilkinson, Denys H.",
    title = "{Analysis of neutron beta decay}",
    doi = "10.1016/0375-9474(82)90051-3",
    journal = "Nucl. Phys. A",
    volume = "377",
    pages = "474--504",
    year = "1982"
}

@article{Sirlin:1967zza,
    author = "Sirlin, A.",
    title = "{General Properties of the Electromagnetic Corrections to the Beta Decay of a Physical Nucleon}",
    doi = "10.1103/PhysRev.164.1767",
    journal = "Phys. Rev.",
    volume = "164",
    pages = "1767--1775",
    year = "1967"
}

@article{Cirigliano:2022hob,
    author = "Cirigliano, Vincenzo and de Vries, Jordy and Hayen, Leendert and Mereghetti, Emanuele and Walker-Loud, Andr{\'e}",
    title = "{Pion-Induced Radiative Corrections to Neutron {\ensuremath{\beta}} Decay}",
    eprint = "2202.10439",
    archivePrefix = "arXiv",
    primaryClass = "nucl-th",
    reportNumber = "LA-UR-21-31960, INT-PUB-22-005, LA-UR-21-31960; INT-PUB-22-005",
    doi = "10.1103/PhysRevLett.129.121801",
    journal = "Phys. Rev. Lett.",
    volume = "129",
    number = "12",
    pages = "121801",
    year = "2022"
}

@article{Ando:2004rk,
    author = "Ando, S. and Fearing, H. W. and Gudkov, Vladimir P. and Kubodera, K. and Myhrer, F. and Nakamura, S. and Sato, T.",
    title = "{Neutron beta decay in effective field theory}",
    eprint = "nucl-th/0402100",
    archivePrefix = "arXiv",
    reportNumber = "TRI-PP-04-05, USC-NT-04-04, USC(NT)-04-04",
    doi = "10.1016/j.physletb.2004.06.037",
    journal = "Phys. Lett. B",
    volume = "595",
    pages = "250--259",
    year = "2004"
}

@article{Cabibbo:1963yz,
    author = "Cabibbo, Nicola",
    title = "{Unitary Symmetry and Leptonic Decays}",
    doi = "10.1103/PhysRevLett.10.531",
    journal = "Phys. Rev. Lett.",
    volume = "10",
    pages = "531--533",
    year = "1963"
}

@article{Kobayashi:1973fv,
    author = "Kobayashi, Makoto and Maskawa, Toshihide",
    title = "{CP Violation in the Renormalizable Theory of Weak Interaction}",
    reportNumber = "KUNS-242",
    doi = "10.1143/PTP.49.652",
    journal = "Prog. Theor. Phys.",
    volume = "49",
    pages = "652--657",
    year = "1973"
}

@article{Abers:1968zz,
    author = "Abers, Ernest S. and Dicus, Duane A. and Norton, Richard E. and Quinn, Helen R.",
    title = "{Radiative Corrections to the Fermi Part of Strangeness-Conserving beta Decay}",
    doi = "10.1103/PhysRev.167.1461",
    journal = "Phys. Rev.",
    volume = "167",
    pages = "1461--1478",
    year = "1968"
}

@article{Jaus:1970tah,
    author = "Jaus, W. and Rasche, G.",
    title = "{Radiative corrections of order $Z\alpha^2$ to $0^+ - 0^+$ $\beta$-transitions}",
    doi = "10.1016/0375-9474(70)90690-1",
    journal = "Nucl. Phys. A",
    volume = "143",
    pages = "202--212",
    year = "1970"
}

@article{Jaus:1986te,
    author = "Jaus, W. and Rasche, G.",
    title = "{Radiative corrections to $0^+ -0^+$ beta transitions}",
    reportNumber = "Print-86-1231 (ZURICH)",
    doi = "10.1103/PhysRevD.35.3420",
    journal = "Phys. Rev. D",
    volume = "35",
    pages = "3420",
    year = "1987"
}

@article{Sirlin:1986cc,
    author = "Sirlin, A. and Zucchini, R.",
    title = "{Accurate Verification of the Conserved Vector Current and Standard Model Predictions}",
    doi = "10.1103/PhysRevLett.57.1994",
    journal = "Phys. Rev. Lett.",
    volume = "57",
    pages = "1994--1997",
    year = "1986"
}

@article{Czarnecki:2004cw,
    author = "Czarnecki, Andrzej and Marciano, William J. and Sirlin, Alberto",
    title = "{Precision measurements and CKM unitarity}",
    eprint = "hep-ph/0406324",
    archivePrefix = "arXiv",
    reportNumber = "ALBERTA-THY-12-04, BNL-HET-04-7, NYU-TH-04-06-25",
    doi = "10.1103/PhysRevD.70.093006",
    journal = "Phys. Rev. D",
    volume = "70",
    pages = "093006",
    year = "2004"
}

@article{Towner:2010zz,
    author = "Towner, I. S. and Hardy, J. C.",
    title = "{The evaluation of V(ud) and its impact on the unitarity of the Cabibbo-Kobayashi-Maskawa quark-mixing matrix}",
    doi = "10.1088/0034-4885/73/4/046301",
    journal = "Rept. Prog. Phys.",
    volume = "73",
    pages = "046301",
    year = "2010"
}

@article{Marciano:2005ec,
    author = "Marciano, William J. and Sirlin, Alberto",
    title = "{Improved calculation of electroweak radiative corrections and the value of V(ud)}",
    eprint = "hep-ph/0510099",
    archivePrefix = "arXiv",
    doi = "10.1103/PhysRevLett.96.032002",
    journal = "Phys. Rev. Lett.",
    volume = "96",
    pages = "032002",
    year = "2006"
}

@article{Wilkinson:1993hxz,
    author = "Wilkinson, Denys H.",
    title = "{Methodology for superallowed Fermi beta decay. 1. Preliminaries and data}",
    reportNumber = "TRI-PP-93-19",
    doi = "10.1016/0168-9002(93)90270-R",
    journal = "Nucl. Instrum. Meth. A",
    volume = "335",
    pages = "172--181",
    year = "1993"
}

@article{Wilkinson:1993fva,
    author = "Wilkinson, Denys H.",
    title = "{Methodology for superallowed Fermi beta decay. 3. Analysis}",
    reportNumber = "TRI-PP-93-21",
    doi = "10.1016/0168-9002(93)90272-J",
    journal = "Nucl. Instrum. Meth. A",
    volume = "335",
    pages = "201--218",
    year = "1993"
}

@article{Sirlin:1977sv,
    author = "Sirlin, A.",
    title = "{Current Algebra Formulation of Radiative Corrections in Gauge Theories and the Universality of the Weak Interactions}",
    reportNumber = "NYU-TR12-77",
    doi = "10.1103/RevModPhys.50.573",
    journal = "Rev. Mod. Phys.",
    volume = "50",
    pages = "573",
    year = "1978",
    note = "[Erratum: Rev.Mod.Phys. 50, 905 (1978)]"
}

@article{Sirlin:1967zz,
    author = "Sirlin, A.",
    title = "{Electromagnetic Corrections, Current Algebra, and the Intermediate Boson}",
    doi = "10.1103/PhysRevLett.19.877",
    journal = "Phys. Rev. Lett.",
    volume = "19",
    pages = "877--880",
    year = "1967"
}

@article{Kinoshita:1958ru,
    author = "Kinoshita, Toichiro and Sirlin, Alberto",
    title = "{Radiative corrections to Fermi interactions}",
    doi = "10.1103/PhysRev.113.1652",
    journal = "Phys. Rev.",
    volume = "113",
    pages = "1652--1660",
    year = "1959"
}

@article{Towner:1992xm,
    author = "Towner, I. S.",
    title = "{The Nuclear structure dependence of radiative corrections in superallowed Fermi beta decay}",
    doi = "10.1016/0375-9474(92)90170-O",
    journal = "Nucl. Phys. A",
    volume = "540",
    pages = "478--500",
    year = "1992"
}

@article{Seng:2018yzq,
    author = "Seng, Chien-Yeah and Gorchtein, Mikhail and Patel, Hiren H. and Ramsey-Musolf, Michael J.",
    title = "{Reduced Hadronic Uncertainty in the Determination of $V_{ud}$}",
    eprint = "1807.10197",
    archivePrefix = "arXiv",
    primaryClass = "hep-ph",
    doi = "10.1103/PhysRevLett.121.241804",
    journal = "Phys. Rev. Lett.",
    volume = "121",
    number = "24",
    pages = "241804",
    year = "2018"
}

@article{Seng:2018qru,
    author = "Seng, Chien Yeah and Gorchtein, Mikhail and Ramsey-Musolf, Michael J.",
    title = "{Dispersive evaluation of the inner radiative correction in neutron and nuclear $\beta$ decay}",
    eprint = "1812.03352",
    archivePrefix = "arXiv",
    primaryClass = "nucl-th",
    reportNumber = "MITP/18-096; ACFI-T18-19",
    doi = "10.1103/PhysRevD.100.013001",
    journal = "Phys. Rev. D",
    volume = "100",
    number = "1",
    pages = "013001",
    year = "2019"
}

@article{Ma:2023kfr,
    author = "Ma, Peng-Xiang and Feng, Xu and Gorchtein, Mikhail and Jin, Lu-Chang and Liu, Keh-Fei and Seng, Chien-Yeah and Wang, Bi-Geng and Zhang, Zhao-Long",
    title = "{Lattice QCD Calculation of Electroweak Box Contributions to Superallowed Nuclear and Neutron Beta Decays}",
    eprint = "2308.16755",
    archivePrefix = "arXiv",
    primaryClass = "hep-lat",
    doi = "10.1103/PhysRevLett.132.191901",
    journal = "Phys. Rev. Lett.",
    volume = "132",
    number = "19",
    pages = "191901",
    year = "2024"
}

@article{Seng:2022cnq,
    author = "Seng, Chien-Yeah and Gorchtein, Mikhail",
    title = "{Dispersive formalism for the nuclear structure correction {\ensuremath{\delta}}$_{\text{NS}}$ to the {\ensuremath{\beta}} decay rate}",
    eprint = "2211.10214",
    archivePrefix = "arXiv",
    primaryClass = "nucl-th",
    doi = "10.1103/PhysRevC.107.035503",
    journal = "Phys. Rev. C",
    volume = "107",
    number = "3",
    pages = "035503",
    year = "2023"
}

@article{Seng:2023cvt,
    author = "Seng, Chien-Yeah and Gorchtein, Mikhail",
    title = "{Toward ab-initio nuclear theory calculations of {\ensuremath{\delta}}$_C$}",
    eprint = "2304.03800",
    archivePrefix = "arXiv",
    primaryClass = "nucl-th",
    doi = "10.1103/PhysRevC.109.044302",
    journal = "Phys. Rev. C",
    volume = "109",
    number = "4",
    pages = "044302",
    year = "2024"
}

@article{Gorchtein:2018fxl,
    author = "Gorchtein, Mikhail",
    title = "{{\ensuremath{\gamma}}W Box Inside Out: Nuclear Polarizabilities Distort the Beta Decay Spectrum}",
    eprint = "1812.04229",
    archivePrefix = "arXiv",
    primaryClass = "nucl-th",
    reportNumber = "MITP/18-096, MITP-18-096",
    doi = "10.1103/PhysRevLett.123.042503",
    journal = "Phys. Rev. Lett.",
    volume = "123",
    number = "4",
    pages = "042503",
    year = "2019"
}

@article{Cirigliano:2023fnz,
    author = "Cirigliano, Vincenzo and Dekens, Wouter and Mereghetti, Emanuele and Tomalak, Oleksandr",
    title = "{Effective field theory for radiative corrections to charged-current processes: Vector coupling}",
    eprint = "2306.03138",
    archivePrefix = "arXiv",
    primaryClass = "hep-ph",
    reportNumber = "LA-UR-22-21034, INT-PUB-23-015",
    doi = "10.1103/PhysRevD.108.053003",
    journal = "Phys. Rev. D",
    volume = "108",
    number = "5",
    pages = "053003",
    year = "2023"
}

@article{Borah:2024ghn,
    author = "Borah, Kaushik and Hill, Richard J. and Plestid, Ryan",
    title = "{Renormalization of beta decay at three loops and beyond}",
    eprint = "2402.13307",
    archivePrefix = "arXiv",
    primaryClass = "hep-ph",
    reportNumber = "FERMILAB-PUB-24-0070-T, CALT-TH-2024-006",
    doi = "10.1103/PhysRevD.109.113007",
    journal = "Phys. Rev. D",
    volume = "109",
    number = "11",
    pages = "113007",
    year = "2024"
}

@article{Sirlin:1981ie,
    author = "Sirlin, A.",
    title = "{Large $m_W$, $m_Z$ Behavior of the $O(\alpha)$ Corrections to Semileptonic Processes Mediated by W}",
    reportNumber = "NYU/TR8/81",
    doi = "10.1016/0550-3213(82)90303-0",
    journal = "Nucl. Phys. B",
    volume = "196",
    pages = "83--92",
    year = "1982"
}

@article{Hill:2019xqk,
    author = "Hill, Richard J. and Tomalak, Oleksandr",
    title = "{On the effective theory of neutrino-electron and neutrino-quark interactions}",
    eprint = "1911.01493",
    archivePrefix = "arXiv",
    primaryClass = "hep-ph",
    reportNumber = "FERMILAB-PUB-19-559-T",
    doi = "10.1016/j.physletb.2020.135466",
    journal = "Phys. Lett. B",
    volume = "805",
    pages = "135466",
    year = "2020"
}

@article{Dekens:2019ept,
    author = "Dekens, Wouter and Stoffer, Peter",
    title = "{Low-energy effective field theory below the electroweak scale: matching at one loop}",
    eprint = "1908.05295",
    archivePrefix = "arXiv",
    primaryClass = "hep-ph",
    doi = "10.1007/JHEP10(2019)197",
    journal = "JHEP",
    volume = "10",
    pages = "197",
    year = "2019",
    note = "[Erratum: JHEP 11, 148 (2022)]"
}

@article{Hardy:2020qwl,
    author = "Hardy, J. C. and Towner, I. S.",
    title = "{Superallowed $0^+ \to 0^+$ nuclear $\beta$ decays: 2020 critical survey, with implications for V$_{ud}$ and CKM unitarity}",
    doi = "10.1103/PhysRevC.102.045501",
    journal = "Phys. Rev. C",
    volume = "102",
    number = "4",
    pages = "045501",
    year = "2020"
}

@article{King:2025fph,
    author = "King, Garrett B. and Carlson, Joseph and Flores, Abraham R. and Gandolfi, Stefano and Mereghetti, Emanuele and Pastore, Saori and Piarulli, Maria and Wiringa, Robert B.",
    title = "{Quantum Monte Carlo calculation of $\delta_{\rm NS}$ in $^{10}$C using an effective field theory approach}",
    eprint = "2509.07310",
    archivePrefix = "arXiv",
    primaryClass = "nucl-th",
    reportNumber = "LA-UR-25-28377",
    month = "9",
    year = "2025"
}

@article{Gennari:2024sbn,
    author = "Gennari, Michael and Drissi, Mehdi and Gorchtein, Mikhail and Navratil, Petr and Seng, Chien-Yeah",
    title = "{Ab~Initio Strategy for Taming the Nuclear-Structure Dependence of Vud Extractions: The $^{10}$C {\textrightarrow} $^{10}$B Superallowed Transition}",
    eprint = "2405.19281",
    archivePrefix = "arXiv",
    primaryClass = "nucl-th",
    doi = "10.1103/PhysRevLett.134.012501",
    journal = "Phys. Rev. Lett.",
    volume = "134",
    number = "1",
    pages = "012501",
    year = "2025"
}

@article{Fermi:1934hr,
    author = "Fermi, E.",
    title = "{An attempt of a theory of beta radiation. 1.}",
    reportNumber = "UCRL-TRANS-726",
    doi = "10.1007/BF01351864",
    journal = "Z. Phys.",
    volume = "88",
    pages = "161--177",
    year = "1934"
}

@article{Jenkins:1990jv,
    author = "Jenkins, Elizabeth Ellen and Manohar, Aneesh V.",
    title = "{Baryon chiral perturbation theory using a heavy fermion Lagrangian}",
    reportNumber = "UCSD-PTH-90-23",
    doi = "10.1016/0370-2693(91)90266-S",
    journal = "Phys. Lett. B",
    volume = "255",
    pages = "558--562",
    year = "1991"
}

@article{Georgi:1990um,
    author = "Georgi, Howard",
    title = "{An Effective Field Theory for Heavy Quarks at Low-energies}",
    reportNumber = "HUTP-90/A007",
    doi = "10.1016/0370-2693(90)91128-X",
    journal = "Phys. Lett. B",
    volume = "240",
    pages = "447--450",
    year = "1990"
}

@article{Czarnecki:2019mwq,
    author = "Czarnecki, Andrzej and Marciano, William J. and Sirlin, Alberto",
    title = "{Radiative Corrections to Neutron and Nuclear Beta Decays Revisited}",
    eprint = "1907.06737",
    archivePrefix = "arXiv",
    primaryClass = "hep-ph",
    reportNumber = "Alberta-Thy-7-19, Alberta Thy 7-19",
    doi = "10.1103/PhysRevD.100.073008",
    journal = "Phys. Rev. D",
    volume = "100",
    number = "7",
    pages = "073008",
    year = "2019"
}

@article{Shiells:2020fqp,
    author = "Shiells, K. and Blunden, P. G. and Melnitchouk, W.",
    title = "{Electroweak axial structure functions and improved extraction of the Vud CKM matrix element}",
    eprint = "2012.01580",
    archivePrefix = "arXiv",
    primaryClass = "hep-ph",
    reportNumber = "JLAB-THY-20-3289",
    doi = "10.1103/PhysRevD.104.033003",
    journal = "Phys. Rev. D",
    volume = "104",
    number = "3",
    pages = "033003",
    year = "2021"
}

@article{Hayen:2020cxh,
    author = "Hayen, Leendert",
    title = "{Standard model $\mathcal{O}(\alpha)$ renormalization of $g_A$ and its impact on new physics searches}",
    eprint = "2010.07262",
    archivePrefix = "arXiv",
    primaryClass = "hep-ph",
    doi = "10.1103/PhysRevD.103.113001",
    journal = "Phys. Rev. D",
    volume = "103",
    number = "11",
    pages = "113001",
    year = "2021"
}

@article{Seng:2020wjq,
    author = "Seng, Chien-Yeah and Feng, Xu and Gorchtein, Mikhail and Jin, Lu-Chang",
    title = "{Joint lattice QCD{\textendash}dispersion theory analysis confirms the quark-mixing top-row unitarity deficit}",
    eprint = "2003.11264",
    archivePrefix = "arXiv",
    primaryClass = "hep-ph",
    doi = "10.1103/PhysRevD.101.111301",
    journal = "Phys. Rev. D",
    volume = "101",
    number = "11",
    pages = "111301",
    year = "2020"
}

@article{Cirigliano:2022yyo,
    author = "Cirigliano, Vincenzo and Crivellin, Andreas and Hoferichter, Martin and Moulson, Matthew",
    title = "{Scrutinizing CKM unitarity with a new measurement of the K{\ensuremath{\mu}}3/K{\ensuremath{\mu}}2 branching fraction}",
    eprint = "2208.11707",
    archivePrefix = "arXiv",
    primaryClass = "hep-ph",
    reportNumber = "INT-PUB-22-024, PSI-PR-22-28, ZU-TH 43/22",
    doi = "10.1016/j.physletb.2023.137748",
    journal = "Phys. Lett. B",
    volume = "838",
    pages = "137748",
    year = "2023"
}

@article{Hayen:2017pwg,
    author = "Hayen, Leendert and Severijns, Nathal and Bodek, Kazimierz and Rozpedzik, Dagmara and Mougeot, Xavier",
    title = "{High precision analytical description of the allowed $\beta$ spectrum shape}",
    eprint = "1709.07530",
    archivePrefix = "arXiv",
    primaryClass = "nucl-th",
    doi = "10.1103/RevModPhys.90.015008",
    journal = "Rev. Mod. Phys.",
    volume = "90",
    number = "1",
    pages = "015008",
    year = "2018"
}

@article{Gasser:1983yg,
    author = "Gasser, J. and Leutwyler, H.",
    title = "{Chiral Perturbation Theory to One Loop}",
    reportNumber = "CERN-TH-3689",
    doi = "10.1016/0003-4916(84)90242-2",
    journal = "Annals Phys.",
    volume = "158",
    pages = "142",
    year = "1984"
}

@article{Henn:2014qga,
    author = "Henn, Johannes M.",
    title = "{Lectures on differential equations for Feynman integrals}",
    eprint = "1412.2296",
    archivePrefix = "arXiv",
    primaryClass = "hep-ph",
    doi = "10.1088/1751-8113/48/15/153001",
    journal = "J. Phys. A",
    volume = "48",
    pages = "153001",
    year = "2015"
}

@article{Remiddi:1997ny,
    author = "Remiddi, Ettore",
    title = "{Differential equations for Feynman graph amplitudes}",
    eprint = "hep-th/9711188",
    archivePrefix = "arXiv",
    reportNumber = "DFUB-97-15, DFUB 97-15",
    doi = "10.1007/BF03185566",
    journal = "Nuovo Cim. A",
    volume = "110",
    pages = "1435--1452",
    year = "1997"
}

@article{Gehrmann:1999as,
    author = "Gehrmann, T. and Remiddi, E.",
    title = "{Differential equations for two-loop four-point functions}",
    eprint = "hep-ph/9912329",
    archivePrefix = "arXiv",
    reportNumber = "TTP-99-49",
    doi = "10.1016/S0550-3213(00)00223-6",
    journal = "Nucl. Phys. B",
    volume = "580",
    pages = "485--518",
    year = "2000"
}

@article{Argeri:2007up,
    author = "Argeri, Mario and Mastrolia, Pierpaolo",
    title = "{Feynman Diagrams and Differential Equations}",
    eprint = "0707.4037",
    archivePrefix = "arXiv",
    primaryClass = "hep-ph",
    reportNumber = "ZU-TH-19-07",
    doi = "10.1142/S0217751X07037147",
    journal = "Int. J. Mod. Phys. A",
    volume = "22",
    pages = "4375--4436",
    year = "2007"
}

@book{Smirnov:2012gma,
    author = "Smirnov, Vladimir A.",
    title = "{Analytic tools for Feynman integrals}",
    doi = "10.1007/978-3-642-34886-0",
    volume = "250",
    year = "2012"
}

@book{Collins:1984xc,
    author = "Collins, John C.",
    title = "{Renormalization : An Introduction to Renormalization, the Renormalization Group and the Operator-Product Expansion}",
    doi = "10.1017/9781009401807",
    isbn = "978-0-521-31177-9, 978-0-511-86739-2, 978-1-009-40180-7, 978-1-009-40176-0, 978-1-009-40179-1",
    publisher = "Cambridge University Press",
    address = "Cambridge",
    series = "Cambridge Monographs on Mathematical Physics",
    volume = "26",
    year = "1984"
}

@book{Manohar:2000dt,
    author = "Manohar, Aneesh V. and Wise, Mark B.",
    title = "{Heavy quark physics}",
    doi = "10.1017/9781009402125",
    isbn = "978-0-521-03757-0, 978-1-009-40212-5",
    volume = "10",
    year = "2000"
}

@article{Frixione:1995ms,
    author = "Frixione, S. and Kunszt, Z. and Signer, A.",
    title = "{Three jet cross-sections to next-to-leading order}",
    eprint = "hep-ph/9512328",
    archivePrefix = "arXiv",
    reportNumber = "SLAC-PUB-7073, SLAC-PUB-95-7073, ETH-TH-95-42",
    doi = "10.1016/0550-3213(96)00110-1",
    journal = "Nucl. Phys. B",
    volume = "467",
    pages = "399--442",
    year = "1996"
}

@article{Alioli:2010xd,
    author = "Alioli, Simone and Nason, Paolo and Oleari, Carlo and Re, Emanuele",
    title = "{A general framework for implementing NLO calculations in shower Monte Carlo programs: the POWHEG BOX}",
    eprint = "1002.2581",
    archivePrefix = "arXiv",
    primaryClass = "hep-ph",
    reportNumber = "DESY-10-018, SFB-CPP-10-22, IPPP-10-11, DCPT-10-22",
    doi = "10.1007/JHEP06(2010)043",
    journal = "JHEP",
    volume = "06",
    pages = "043",
    year = "2010"
}

@article{Moretti:2025qxt,
    author = {Moretti, Francesco and Gorbahn, Martin and J{\"a}ger, Sebastian},
    title = "{Beyond Leading Logarithms in $g_V$: The Semileptonic Weak Hamiltonian at $\mathcal{O}(\alpha\,\alpha_s^2)$}",
    eprint = "2510.27648",
    archivePrefix = "arXiv",
    primaryClass = "hep-ph",
    reportNumber = "P3H-25-087, TTP25-041",
    month = "10",
    year = "2025"
}

@article{Cirigliano:2024rfk,
    author = "Cirigliano, Vincenzo and Dekens, Wouter and de Vries, Jordy and Gandolfi, Stefano and Hoferichter, Martin and Mereghetti, Emanuele",
    title = "{Radiative Corrections to Superallowed {\ensuremath{\beta}} Decays in Effective Field Theory}",
    eprint = "2405.18469",
    archivePrefix = "arXiv",
    primaryClass = "hep-ph",
    reportNumber = "INT-PUB-24-020, LA-UR-24-25162",
    doi = "10.1103/PhysRevLett.133.211801",
    journal = "Phys. Rev. Lett.",
    volume = "133",
    number = "21",
    pages = "211801",
    year = "2024"
}

@article{Tarrach:1980up,
    author = "Tarrach, R.",
    title = "{The Pole Mass in Perturbative QCD}",
    reportNumber = "CPT-80/P-1215",
    doi = "10.1016/0550-3213(81)90140-1",
    journal = "Nucl. Phys. B",
    volume = "183",
    pages = "384--396",
    year = "1981"
}

@article{Cao:2025lrw,
    author = "Cao, Zehua and Hill, Richard J. and Plestid, Ryan and Vander Griend, Peter",
    title = "{Factorization and resummation of QED radiative corrections for neutron beta decay}",
    eprint = "2508.05741",
    archivePrefix = "arXiv",
    primaryClass = "hep-ph",
    reportNumber = "CALT-TH-2025-017, FERMILAB-PUB-25-0375-T",
    month = "8",
    year = "2025"
}

@article{VanderGriend:2025mdc,
    author = "Vander Griend, Peter and Cao, Zehua and Hill, Richard J. and Plestid, Ryan",
    title = "{The Fermi function and the neutron's lifetime}",
    eprint = "2501.17916",
    archivePrefix = "arXiv",
    primaryClass = "hep-ph",
    reportNumber = "CALT-TH-2025-001, FERMILAB-PUB-25-0028-T",
    doi = "10.1016/j.physletb.2025.139678",
    journal = "Phys. Lett. B",
    volume = "868",
    pages = "139678",
    year = "2025"
}

@article{Alarcon:2023gfu,
    author = "Alarcon, R. and others",
    title = "{Fundamental Neutron Physics: a White Paper on Progress and Prospects in the US}",
    eprint = "2308.09059",
    archivePrefix = "arXiv",
    primaryClass = "nucl-ex",
    month = "8",
    year = "2023"
}

@article{Plestid:2024eib,
    author = "Plestid, Ryan",
    title = "{Generalized eikonal identities for charged currents}",
    eprint = "2402.14769",
    archivePrefix = "arXiv",
    primaryClass = "hep-ph",
    reportNumber = "CALT-TH/2024-007",
    doi = "10.1007/JHEP07(2024)216",
    journal = "JHEP",
    volume = "07",
    pages = "216",
    year = "2024"
}

@article{Plestid:toappear,
    author = "Cao, Zehua and Hill, Richard J. and Plestid, Ryan and Vander Griend, Peter",
    title = "{The $Z\alpha^2$ correction to superallowed beta decays in effective field theory and implications for $|V_{ud}|$}",
    eprint = "2511.05446",
    archivePrefix = "arXiv",
    primaryClass = "hep-ph",
    reportNumber = "FERMILAB-PUB-25-0773-T, CALT-TH-2025-032, CERN-TH-2025-20",
    month = "11",
    year = "2025"
}

@article{Hocker:2001xe,
    author = "Hocker, Andreas and Lacker, H. and Laplace, S. and Le Diberder, F.",
    title = "{A New approach to a global fit of the CKM matrix}",
    eprint = "hep-ph/0104062",
    archivePrefix = "arXiv",
    reportNumber = "LAL-01-06",
    doi = "10.1007/s100520100729",
    journal = "Eur. Phys. J. C",
    volume = "21",
    pages = "225--259",
    year = "2001"
}

@article{Seng:2023ynd,
    author = "Seng, Chien-Yeah",
    title = "{Pseudo-neutrino versus recoil formalism for 4-body phase space and applications to nuclear decay}",
    eprint = "2312.08630",
    archivePrefix = "arXiv",
    primaryClass = "nucl-th",
    doi = "10.1103/PhysRevC.109.035501",
    journal = "Phys. Rev. C",
    volume = "109",
    number = "3",
    pages = "035501",
    year = "2024"
}

@article{UTfit:2005ras,
    author = "Bona, Marcella and others",
    collaboration = "UTfit",
    title = "{The 2004 UTfit collaboration report on the status of the unitarity triangle in the standard model}",
    eprint = "hep-ph/0501199",
    archivePrefix = "arXiv",
    doi = "10.1088/1126-6708/2005/07/028",
    journal = "JHEP",
    volume = "07",
    pages = "028",
    year = "2005"
}

@article{ParticleDataGroup:2024cfk,
    author = "Navas, S. and others",
    collaboration = "Particle Data Group",
    title = "{Review of particle physics}",
    doi = "10.1103/PhysRevD.110.030001",
    journal = "Phys. Rev. D",
    volume = "110",
    number = "3",
    pages = "030001",
    year = "2024"
}

@misc{fermat,
  title = {Computer Algebra System Fermat},
  howpublished = {\url{https://home.bway.net/lewis}},
  author = {R. H. Lewis}
}

@article{firefly1,
    author = "Klappert, Jonas and Klein, Sven Yannick and Lange, Fabian",
    title = "{Interpolation of dense and sparse rational functions and other improvements in FireFly}",
    eprint = "2004.01463",
    archivePrefix = "arXiv",
    primaryClass = "cs.MS",
    reportNumber = "TTK-20-07, P3H-20-010",
    doi = "10.1016/j.cpc.2021.107968",
    journal = "Comput. Phys. Commun.",
    volume = "264",
    pages = "107968",
    year = "2021"
}

@article{firefly2,
    author = "Klappert, Jonas and Lange, Fabian",
    title = "{Reconstructing rational functions with FireFly}",
    eprint = "1904.00009",
    archivePrefix = "arXiv",
    primaryClass = "cs.SC",
    reportNumber = "TTK-19-11, P3H-19-007",
    doi = "10.1016/j.cpc.2019.106951",
    journal = "Comput. Phys. Commun.",
    volume = "247",
    pages = "106951",
    year = "2020"
}

\end{document}